\documentclass[onecolumn,aps,prd,amsmath,nofootinbib,superscriptaddress]{revtex4-1}

\usepackage{amssymb}

\usepackage{CJK}                     
\usepackage[dvips]{graphicx}
\usepackage{bm}                      
\usepackage{times}
\usepackage{dcolumn}                 
\usepackage{array}
\usepackage{graphicx}
\usepackage{dcolumn}
\usepackage{bm}
\usepackage[usenames]{color}
\usepackage{epstopdf}
\usepackage{epsfig}
\usepackage{float}
\usepackage{subfigure}
\usepackage{multirow}
\usepackage{slashed}
\usepackage{diagbox}
\usepackage{cancel}
\usepackage{graphics}

\newcommand{\nc}{\newcommand}       
\nc{\vc}[1] {\mbox{\boldmath $#1$}} 
\nc{\del}       {\partial}              
\nc{\bra}       {\langle}               
\nc{\ket}       {\rangle}               
\nc{\bras}[1]   {\langle #1|}           
\nc{\kets}[1]   {|#1\rangle}            
\nc{\mapleft}[1]{           
 \smash{\mathop{\,          %
  \hbox to 1.5cm{\rightarrowfill}\, }\limits_{#1}}}
\nc{\nn}      {\\\nonumber} \nc{\vs}      {\vspace{-0.275cm}}
\nc{\fra}    {\frac{1}{2}}
\nc{\mb}        {\mathbf}

\usepackage{color}

\usepackage{hyperref}
\hypersetup{
  colorlinks=true,        
  linkcolor=blue,         
  citecolor=cyan,         
  urlcolor=cyan            
}
\usepackage{natbib,times}
\newcommand{\cf}{cf.~}
\newcommand{\ie}{i.e.,~}
\newcommand{\eg}{e.g.,~}

\begin{document}

\title{Tidal deformability and gravitational-wave phase evolution of magnetised compact-star
  binaries}

\author{Zhenyu Zhu}
\affiliation{Institut f{\"u}r Theoretische Physik, Max-von-Laue-Stra{\ss}e 1, 60438 Frankfurt, Germany}
\affiliation{Department of Astronomy, Xiamen University, Xiamen 361005, China}

\author{Ang Li}
\affiliation{Department of Astronomy, Xiamen University, Xiamen 361005, China}
\author{Luciano Rezzolla} \affiliation{Institut f{\"u}r Theoretische
  Physik, Max-von-Laue-Stra{\ss}e 1, 60438 Frankfurt, Germany}
\affiliation{School of Mathematics, Trinity College, Dublin 2, Ireland}
\affiliation{Helmholtz Research Academy Hesse for FAIR,
  Max-von-Laue-Str. 12, 60438 Frankfurt, Germany}

\date{\today}

\begin{abstract}
The evolution of the gravitational-wave phase in the signal produced by
inspiralling binaries of compact stars is modified by the nonzero
deformability of the two stars. Hence, the measurement of these
corrections has the potential of providing important information on the
equation of state of nuclear matter. Extensive work has been carried out
over the last decade to quantify these corrections, but it has so far
been restricted to stars with zero intrinsic magnetic fields. While the
corrections introduced by the magnetic tension and magnetic pressure are
expected to be subdominant, it is nevertheless useful to determine the
precise conditions under which these corrections become important. To
address this question, we have carried out a second-order perturbative
analysis of the tidal deformability of magnetised compact stars under a
variety of magnetic-field strengths and equations of state describing
either neutron stars or quark stars. Overall, we find that magnetically
induced corrections to the tidal deformability will produce changes in
the gravitational-wave phase evolution that are unlikely to be detected
for a realistic magnetic field \ie $B\sim 10^{10} - 10^{12}\,{\rm G}$. At
the same time, if the magnetic field is unrealistically large, \ie $B\sim
10^{16}\,{\rm G}$, these corrections would produce a sizeable
contribution to the phase evolution, especially for quark stars. In the
latter case, and if the neglected higher-order terms will remain
  negligible also for very high magnetic fields, the induced phase
differences would represent a unique tool to measure the properties of
the magnetic fields, providing information that is otherwise hard to
quantify.
\end{abstract}

\maketitle
\
\section{Introduction}
\label{sec:introduction}

The detection of the binary neutron-star merger GW170817 from the
LIGO-Virgo Scientific Collaboration \cite{Abbott2017_etal} has marked the
first milestone in multimessenger gravitational-wave (GW) astronomy.
This multimessenger observation alone has helped set tighter constraints
on important properties of neutron stars, such as maximum mass and radii
(see \cite{Margalit2017, Bauswein2017b, Rezzolla2017, Ruiz2017,
  Annala2017, Radice2017b, Most2018, Tews2018a, De2018, Abbott2018b,
  Shibata2019, Koeppel2019}, for an incomplete list).
This event and its constraints have also lead to the exploration of
equations of state (EOSs) for nuclear matter that are not purely
hadronic, such as in the scenarios of hybrid (twin) stars (see, \eg
Refs. \cite{Fattoyev2017, Paschalidis2017, Burgio2018, Montana2018,
  Gomes2018, Li2018b, Li2020}), strange quark stars \cite{Zhou2017}, and
even those scenarios in which a phase transition to quark matter takes
place after the merger \cite{Most2018b, Bauswein2019, Weih2020}.

Some of the most stringent constraints on the EOS coming from
GW170817 are based on the measurement of the tidal deformability, which
is defined as the ratio of the induced multipole moment of a star over
the inducing tidal field from its companion. The dominant contribution to
the tidal deformability comes from the \textit{``even-parity''} (or
\textit{gravitoelectric} or \textit{mass})\footnote{Gravitomagnetic and
  gravitoelectric moments are sometimes referred to as ``electric'' and
  ``magnetic'' \cite{Pani15a}, but this can be confusing when intrinsic
  magnetic fields are taken into account, such as those considered in
  this paper. To avoid a possible confusion, we will not use here the
  nomenclature gravitoelectric/gravitomagnetic and distinguish the
  moments according to their parity (\ie odd and even).}  quadrupole
term, which starts to impact the phase of the GW signal emitted in a
binary at the fifth post-Newtonian (5PN) order. The changes in the phase
evolution become particularly significant in the high-frequency region of
the signal, as the stars are about to merge, as discussed in detail in
Ref. \cite{Harry2018}. The even-parity quadrupolar tidal deformability
$\lambda$ is the ratio between the mass-quadrupole moment of the star,
$\mathcal{Q}_{ij}$, and the quadrupolar tidal field, $\mathcal{E}_{ij}$,
and a first discussion on how to compute it was presented in
Refs. \cite{Flanagan2008, Hinderer08, Hinderer09}. Beyond the leading 5PN
order, higher-orders contributions to the waveform have also been
explored in the literature. In particular, the next-leading-order (6PN)
of the even-parity tidal deformability was computed by
Ref. \cite{Vines:2010ca}, while the \textit{``odd-parity''} (or
\textit{gravitomagnetic} or \textit{mass-current}) tidal deformability
$\sigma$ was computed independently by Damour and Nagar
\cite{Damour:2009} and by Binnington and Poisson
\cite{Binnington:2009bb}, obtaining two master equations that are not
equivalent. Subsequently, Landry and Poisson \cite{Landry2015} have shown
that the odd-parity tidal deformability actually depends on the
assumption made on the properties of the fluids, so that assuming a
static equilibrium or an irrotational flow leads to different
results. Theses ambiguities in the odd-parity tidal deformability were
studied and clarified in Ref. \cite{Pani2018}, where it was shown that
the odd-parity tidal deformabilities computed in Refs. \cite{Damour:2009,
  Landry2015} are equivalent and both are based on irrotational
configurations, whereas the corresponding results from
\cite{Binnington:2009bb} assume a strict static background configuration
and are therefore less realistic (this was concluded already in
Ref. \cite{Landry2015}).

The impact of the odd-parity tidal deformability on the GW phase
evolution was first explored by Yagi \cite{Yagi2014c}, and further
extended in \cite{JimenezForteza2018}, where it was also applied to the
analysis of the signal from GW170817. In general, the corrections to the
phase evolution of odd-parity tidal deformabilities appear at one
post-Newtonian order higher than to the corresponding even-parity ones,
\ie the corrections to the phase evolution from the even- and odd-parity
tidal deformabilities appear at 5PN and 6PN, respectively. A different
behaviour is seen for the GW amplitudes, where the corrections to the
mode amplitudes from the even- and odd-parity tidal deformabilities
appear at 6PN and 5PN, respectively \cite{Batoul2018}. On the hand, for
some modes, \eg $h_{21}$ or $h_{32}$ the contributions start at the same
leading post-Newtonian order, \ie 5PN \cite{Batoul2018}.

The presence of spin angular momentum in the stars also impacts the
calculation of the GW phase of spinning and tidally deformed stars, with
the spin-tidal coupling appearing at 6.5PN for both the even- and the
odd-parity tidal deformabilities \cite{Abdelsalhin2018a,
  JimenezForteza2018}. In particular, the spin angular momentum gives
rise to the coupling between different multipole moments. In the
nonspinning case, the even- and odd-parity quadrupolar tidal fields could
only result in even- and odd-parity quadrupole moments, \ie
\begin{eqnarray}
  \label{eq:moment}
  & & \mathcal{Q}_{ij} = -\lambda_2\mathcal{E}_{ij}\,, \\
  \label{eq:mass-current}
  & & \mathcal{S}_{ij} = -\sigma_2\mathcal{B}_{ij}\,,
\end{eqnarray}
where $\mathcal{Q}_{ij}$ and $\mathcal{S}_{ij}$ denote the even- and
odd-parity (inducing) quadrupolar tidal fields, while $\mathcal{E}_{ij}$
and $\mathcal{B}_{ij}$ are the are corresponding even- and odd-parity
(induced) quadrupole moments. Expressions \eqref{eq:moment} and
\eqref{eq:mass-current} essentially define $\lambda_2$ and $\sigma_2$ as
the ratios between the inducing quadrupolar tidal fields and the
corresponding quadrupolar deformations for the two different parities. If
the stars are spinning, however, the coupling between quadrupole and
octupole moment leads quadrupole-octupole tidal deformabilites
\begin{eqnarray}
& & \mathcal{Q}_{ij} = -\lambda_2\mathcal{E}_{ij} + \lambda_{23}J^{k}\mathcal{E}_{ijk}\,, 
\label{eq:espinmoment}\\
& & \mathcal{S}_{ij} = -\sigma_2\mathcal{B}_{ij} + \sigma_{23}J^{k}\mathcal{B}_{ijk}\,,
\label{eq:ospinmoment}
\end{eqnarray}
where $\mathcal{E}_{ijk}$ and $\mathcal{B}_{ijk}$ are the even- and
odd-parity octupole moments, $J^{k}$ is the spin vector of the star and
$\lambda_{23}$ and $\sigma_{23}$ are respectively the quadrupole-octupole
even- and odd-parity tidal deformabilites. In turn, these deformabilities
lead to a 6.5PN contribution to the GW phase \cite{Abdelsalhin2018a,
  JimenezForteza2018}. While the approach delineated above has been
widely used so far, it has been recently pointed out that it may actually
be flawed \cite{Poisson2020b}. Since we are not considering here a
velocity field in the star, the considerations raised in
Ref. \cite{Poisson2020b} do not affect us directly, but for when we
compare the results of our analysis with the tidal corrections induced by
spin in Sec. \ref{sec:IIIC}.

Oscillation modes in the star could also contribute to the waveform
and phase evolution since they generate a time-varying quadrupolar
moment. The excitation of different oscillation modes in binary system
and its impact on the GW signal and phase evolution have been discussed
in recent work
\cite{Flanagan2007,Steinhoff2016,Hinderer2016,Poisson2020,Ma2020}.
Finally, the effects of elastic crusts on tidal deformability and on the
GW signal are also discussed in Refs. \cite{Pereira2020,Gittins2020},
where it is concluded that elastic crusts are unlikely to generate a
noticeable impact.

We are here also concerned with high-order corrections to the tidal
deformability that are however introduced by the presence of an intrinsic
magnetic field in the stars and should therefore not be confused with the
gravitomagnetic corrections to the tidal deformability discussed
above. At the order considered here, the magnetic field induces
correction only to the even-parity quadrupole moment and we assume that
it does not lead to coupling of different multipole moments. However,
because these represent a correction to the standard unmagnetised,
nonspinning tidal deformability, we are forced to performed an analysis
which includes second-order perturbations. In this way, we are able to
compute the magnetic-field induced changes to the tidal deformability and
to assess their impact on the evolution of the GW phase for different
strengths of the magnetic field and for different EOSs, including those
that describe quark stars. In this way, we find that for realistic
magnetic fields of the order of $10^{12}\,{\rm G}$, the effect on the
phase evolution is too small to be measurable by present and advanced GW
detectors (this point was already explored in numerical simulations
\cite{Giacomazzo:2009mp}). At the same time, these corrections could be
important for third-generation GW detectors such as the Einstein
Telescope (ET) \cite{Punturo:2010} or Cosmic Explorer (CE)
\cite{Abbott17}, or even for advanced detectors in the unlikely scenario
in which one of the stars has magnetic fields of the order of
$10^{16}\,{\rm G}$.

The plan of the paper is as follows. In Sec. \ref{sec:formalism} we
introduce the formalism adopted for the background metric and fluid
variables, for the magnetic-field configuration, the tidal deformability,
and the modifications to the tidal deformability resulting from the
presence of a magnetic field. Our results of tidal-deformability
modifications and their impact on the evolution of the GW phase are
presented in Sec. \ref{sec:results}. Finally, we summarises our findings
in Sec. \ref{sec:summary}. Appendix \ref{sec:appendix_A} provides details
on derivation of some of the equations presented in the main text and
explicit expressions for some of the lengthy source functions.

\section{Mathematical setup}
\label{sec:formalism}

\subsection{Background solution}

At the order considered here, both the magnetic field and the tidal field
are treated as perturbations on a static spherically symmetric spacetime
with background $\boldsymbol{\mathring{g}}$ whose line element can be
written generically as 
\begin{equation}
  \label{eq:g0_munu}
  ds^2 = \mathring{g}_{\mu\nu} dx^{\mu} dx^{\nu} = -e^\nu dt^2 +
  e^\lambda dr^2 + r^2 d\theta^2 + r^2\sin^2\theta\, d\phi^2\,,
\end{equation}  

The metric functions
$\nu$ and $\lambda$ can be obtained by solving the standard
Tolmann-Oppenheimer-Volkov (TOV) equations
\begin{eqnarray}
& & m^\prime = 4\pi r^2e\,, \label{eq:tov1} \\
& & p^\prime = -(e+p)\frac{m+4\pi r^3 p}{r(r - 2m)}\,, \label{eq:tov2}\\
& & \nu^\prime = -\frac{2}{e+p}p^\prime\,, \label{eq:tov3}
\end{eqnarray}
where $e$ and $p$ are, respectively, the energy density and the pressure,
$m(r):=r(1-e^{-\lambda})/2$ is the gravitational mass within the radius
$r$ and a prime $'$ is used to denote a total derivative in the radial
direction. Once the EOS $p = p(e)$ and the central pressure are
specified, the solutions can be obtained by integrating the TOV equations
\eqref{eq:tov1}--\eqref{eq:tov3} from the center up to the surface of the
star (note that $m(0)=0$). The boundary conditions to be specified at the
stellar surface are $m(R)=M$, $p(R)=0$, and $\nu=\ln(1-2M/R)$, where $M$
and $R$ are the stellar mass and radius.

\subsection{First-order magnetic-field perturbations}
\label{sec:fomf}

The magnetic field is assumed to be axially symmetric and purely poloidal
(\ie any meridional electric current is assumed to be zero) \cite{Konno1999, 
Ioka04, Colaiuda2008}. The perturbed metric can then be written as
\begin{equation}
  \label{eq:gmunu_hBmunu}
  g_{\mu \nu} = \mathring{g}_{\mu \nu} + h^{\rm B}_{\mu \nu}\,,
\end{equation}
where the perturbations of the metric resulting from the presence of a
magnetic field can be expanded in terms of spherical-harmonic functions
(since the magnetic field is dipolar, it is sufficient to consider only
the lower-order harmonics, \ie $\ell=0=m$ and $\ell=2, m=0$) and written
as\footnote{Hereafter, we will use an upper index ``${\rm B}$'' to denote
  first-order perturbative quantities associated to the magnetic field of
  strength $B$. Note that although at first order, these perturbative
  quantities are $\mathcal{O}(B^2)$.}
\begin{widetext}
\begin{eqnarray}
h_{\mu\nu}^{\rm B} & = & 
2\begin{pmatrix}
-e^\nu (h_0^{\rm B} + h_2^{\rm B}P_2(\cos\theta) ) & 0 & 0 & 0 \\
0 & e^{2\lambda} \left({m_0^{\rm B}} + {m_2^{\rm B}} P_2(\cos\theta)\right)/{r}  & 0 & 0 \\
0 & 0 & r^2 k_2^{\rm B}P_2(\cos\theta) & 0 \\
0 & 0 & 0 & r^2 k_2^{\rm B} \sin^2\theta P_2(\cos\theta) \\
\end{pmatrix}
\,. \label{eq:permag} \\ \nonumber 
\end{eqnarray}
\end{widetext}
Here, the metric functions $h_0^{\rm B}(r)$, $h_2^{\rm B}(r)$, $m_0^{\rm
  B}(r)$, $m_2^{\rm B}(r)$ and $k_2^{\rm B}(r)$ will be provided via the
solution of Einstein equations, while $P_2(\cos\theta)$ is the Legendre
polynomial of second order. Assuming that the electrical conductivity in
the star is infinite, \ie ideal magnetohydrodynamic (MHD) limit, the MHD
equations can be written as conservation equations for the
energy-momentum tensor $T^{\mu\nu}$, together with the Maxwell equations
for the Faraday tensor, $F^{\mu\nu}$, and the electromagnetic current,
$J^{\mu}$, \ie
\begin{equation}
\nabla_\nu T^{\mu\nu} = 0\,,  \qquad \qquad \nabla_{\nu}F^{\mu\nu} =
J^\mu\,. \label{eq:mhd}
\end{equation}
The system is then closed by the Einstein equations
\begin{equation}
G_{\mu\nu} = 8\pi T_{\mu\nu}, \label{eq:einstein}
\end{equation}
where $G_{\mu\nu}$ is Einstein tensor. The resulting system of
perturbation equation are then given by~\cite{Konno1999,Colaiuda2008}
%
\begin{eqnarray}
&& a_1^{\prime\prime} + \frac{\nu^\prime-\lambda^\prime}{2}a_1^\prime - \frac{2 e^{\lambda}}{r^2}a_1 =  4\pi(e+p)r^2 e^{\lambda} c_0\,,  \label{eq:eqm11}\\
&& h_2^{\rm B} + \frac{m_2^{\rm B}}{r}e^\lambda = \frac{2}{3}e^{-\lambda}(a_1^{\prime})^{2}\,, \label{eq:eqm12}\\
&& (h_2^{\rm B})^{\prime} + \frac{4e^\lambda}{\nu^\prime r^2}y_2^{\rm B} + \left[\nu^\prime - \frac{8\pi e^\lambda}{\nu^\prime}(e+p) + \frac{2}{r^2 \nu^\prime}(e^\lambda-1) \right]h_2^{\rm B} = \frac{\nu^\prime}{3}e^{-\lambda}(a_1^{\prime})^{2} + \frac{4}{3r^2}a_1 a_1^\prime - \frac{16\pi c_0 e^\lambda}{3\nu^\prime} a_1 (e+p)\,, \label{eq:eqm21}\\
&& (y_2^{\rm B})^{\prime} + \nu^\prime h_2^{\rm B} =
\frac{\nu^\prime}{2}e^{-\lambda}(a_1^{\prime})^{2} - \frac{4\pi
  r^2}{3}c_0\left(a_1^\prime + \frac{2}{r}a_1 \right)(e+p) +
\frac{1}{3}\left[\frac{e^{-\lambda}}{r}(\nu^\prime + \lambda^\prime +
  \frac{2}{r}) - \frac{2}{r^2} \right]a_1 a_1^\prime\,, \label{eq:eqm22} \\
&& (m_0^{\rm B})^{\prime} = 4\pi r^2 (e+p)\frac{1}{c_s^2} P_0 + \frac{e^{-\lambda}}{3}(a_1^{\prime})^{2} + \frac{2}{3r^2}a_1^2, \label{eq:eqm31}\\
&& P_0^\prime = -\left(8\pi p + \frac{1}{r^2} \right)e^{2\lambda} m_0^{\rm B} - 4\pi r e^\lambda (e+p) P_0 -\frac{1}{3r}(a_1^{\prime})^{2} + \frac{2}{3r^3}e^\lambda a_1^2 - \frac{2}{3}c_0 a_1^\prime\,. \label{eq:eqm32}
\end{eqnarray}
%
The functions $y_2^{\rm B}(r)$ and $P_0(r)$ are shorthands introduced to
keep equations compact and are defined as
\begin{eqnarray}
& & y_2^{\rm B} := h_2^{\rm B} + k_2^{\rm B} - \frac{e^{-\lambda}}{6}(a_1^{\prime})^{2} -
  \frac{2e^{-\lambda}}{3r}a_1a_1^\prime -
  \frac{2}{3r^2}a_1^2  \,, \label{eq:y21}
  \\
& &  P_0 := \frac{p^{\rm B}}{e+p}\,, \label{eq:y22}
\end{eqnarray}
where $p^{\rm B}$ is the magnetic-pressure perturbation, while $a_1(r)$
is a function related to the strength of magnetic field. In this way, the
poloidal covariant components of the magnetic field in a locally inertial
frame carried by static observers \cite{Rezzolla2001,Rezzolla2004} can be
written as
\begin{eqnarray}
& & B_r = -\frac{2a_1}{r^2}\cos\theta\,, \label{eq:mag1}\\
& & B_\theta = \frac{e^{-\lambda/2}a_1^\prime}{r}\sin\theta\,. \label{eq:mag2}
\end{eqnarray}
The function $P_0$ can also be used to correlate the metric functions
through the following equation
\begin{eqnarray}
& & P_0 + h_0^{\rm B} - \frac{2}{3}c_0 a_1 = c_1\,, \label{eq:dph0}
\end{eqnarray}
which is derived from the MHD equations~\eqref{eq:mhd}, and where $c_0$
and $c_1$ are two integration constants that can be determined using the
boundary conditions.

Two important remarks are worth making. First, the metric functions $\nu$
and $\lambda$, as well as the fluid quantities $p$ and $e$ appearing in
Eqs. \eqref{eq:eqm11}--\eqref{eq:eqm32}, are those of the background
spacetime. However, the fluid structure of the star is modified by the
presence of a magnetic field in terms of the corrections to the metric
(\eg to the function $m^{\rm B}_0$) and to the pressure (\eg with the
inclusion of the magnetic pressure $P_0$). Second, although the
perturbation is only at first order in the magnetic field, it is
proportional to the square of the magnetic-field strength, since both
$m^{\rm B}_0$ and $P_0$ are proportional to $a^2_1 \propto B^2$.

Before solving Eqs.  \eqref{eq:eqm11}--\eqref{eq:eqm32}, it is useful to
recall the required behavior at the origin. In particular, when $r
\rightarrow 0$, it is possible to derive that the functions below have to
behave as
\begin{eqnarray}
& & a_1(r) \rightarrow \alpha_0 r^2, \qquad h_2^{\rm B}(r) \rightarrow A_h r^2,
  \qquad y_2^{\rm B}(r) \rightarrow A_y r^4\,  \label{eq:inim1} \\
& & m_0^{\rm B}(r) \rightarrow \frac{2\alpha_0^2}{3}  r^3, \qquad P_0(r)
  \rightarrow -\frac{2(\alpha_0^2-c_0\alpha_0)}{3}  r^2\, \label{eq:inim2}
\end{eqnarray}
where
\begin{eqnarray}
  A_y := \left(-2\pi A_h + \frac{16}{3}\pi\alpha_0^2\right)
  \left(p_c + \frac{e_c}{3}\right) -
\frac{4\pi}{3}\alpha_0 c_0 (p_c+e_c)\,.
\label{eq:inim3}
\end{eqnarray}
Here, $A_h$ and $\alpha_0$ are constants and will represent the
initial conditions for the integration, while $p_c$ and $e_c$ denote the
pressure and energy density at the center of star.

Note that Eqs.~\eqref{eq:eqm11}--\eqref{eq:eqm32} refer to the stellar
interior where, $e$ and $p$ are obviously nonzero; the corresponding
exterior equations are identical but with vanishing energy and
pressure. Omitting them here for compactness, we just report the explicit
solution; in particular, for the magnetic field we have
\cite{Rezzolla2001, Zanotti02mn}
\begin{eqnarray}
& & a_1 = -\frac{3\mu}{8M^3}r^2\left[\ln\left(1-\frac{2M}{r}\right) +
    \frac{2M}{r} + \frac{2M^2}{r^2} \right]\,, \label{eq:a1}
\end{eqnarray}
where $\mu$ is the magnetic dipole moment. In practice, we match the
interior and exterior expressions for $a_1$ by requiring it is continuous
and with continuous derivative at the stellar surface. Similarly, the
integration constants $c_0$, $\mu$ and $\alpha_0$ can be determined once
the magnetic-field strength at the pole, $B$, is fixed. Finally, the
exterior solutions for the relevant metric functions are given by (see
Ref.~\cite{Konno1999, Ioka04, Colaiuda2008} for details)
\begin{widetext}
\begin{eqnarray}
h_2^{\rm B} & = & K^{\rm B} Q_2^2(z) + \hat{h}_2^{\rm B}(z)\,, \label{eq:solm1}\\
y_2^{\rm B} & = & -\frac{2K^{\rm B}}{\sqrt{z^2-1}} Q_2^1(z) + \hat{y}_2^{\rm B}(z) - \frac{e^{-\lambda}}{6}(a_1^{\prime})^{2} - \frac{2e^{-\lambda}}{3r}a_1^\prime a_1 - \frac{2}{3r^2}a_1^2\,, \label{eq:solm2} \\
m_0^{\rm B} & = & \frac{3\mu^2}{8M^5}(r^2-Mr-M^2)\ln\left(1-\frac{2M}{r}\right) + \frac{3\mu^2}{32M^6}r^2(r-2M)\ln\left(1-\frac{2M}{r}\right)^2 + \frac{3\mu^2}{8M^4r}(r^2-M^2) + c_2\,, \label{eq:solm3} \\
h_0^{\rm B} & = & \frac{3\mu^2}{8M^5}\frac{(r-M)(r-3M)}{r-2M}\ln\left(1-\frac{2M}{r}\right) + \frac{3\mu^2}{32M^6}r^2\ln\left(1-\frac{2M}{r}\right)^2 \nonumber \\
& & - \frac{c_2}{r-2M} - \frac{3\mu^2}{8M^3}\frac{4r-M}{r(r-2M)} + \frac{3\mu^2}{8M^4}\,. \label{eq:solm4}
\end{eqnarray}
\end{widetext}
Here, $Q_2^1$ and $Q_2^2$ are the associated Legendre functions of second
kind, $z := r/M - 1$, while the functions $\hat{h}_2^{\rm B}(z)$ and
$\hat{y}_2^{\rm B}(z)$ are defined as
\begin{widetext}
\begin{eqnarray}
\hat{y}_2^{\rm B}(z) & := & \frac{3\mu^2}{8M^4}\frac{7z^2-4}{z^2-1} + \frac{3\mu^2}{16M^4}\frac{z(11z^2-7)}{z^2-1}\ln\left(\frac{z-1}{z+1}\right) + \frac{3\mu^2}{16M^4}(2z^2+1)\left[\ln\left(\frac{z-1}{z+1}\right)\right]^2\,, \label{eq:solm21} \\
%
\hat{h}_2^{\rm B}(z) & := & -\frac{3\mu^2}{16M^4}\left\{\left(3z -
\frac{4z^2+2z}{z^2-1}\right) -
(z^2-1)\left[\ln\left(\frac{z-1}{z+1} \right)\right]^2
+ \frac{1}{2}\left(3z^2-8z-3-\frac{8}{z^2-1}
\right)\ln\left(\frac{z-1}{z+1} \right)\right\}\,, 
\label{eq:solm22}
\end{eqnarray}
\end{widetext}
where the integration constants $K^{\rm B}$ and $c_2$ are also determined
by the boundary conditions. With the exterior solution given by
Eqs.~\eqref{eq:solm1}-- \eqref{eq:solm4}, and with the initial conditions
Eqs.~\eqref{eq:inim1}--\eqref{eq:inim2}, the complete set of the
first-order magnetic-field perturbative equations
\eqref{eq:eqm11}--\eqref{eq:eqm32} can then be solved numerically.

Note that the magnetic field will introduce a deformation in the star and
hence a magnetically induced quadrupolar moment. Such an ellipticity and
quadrupolar moment can be computed as~\cite{Konno1999, Colaiuda2008}
\begin{eqnarray}
e^{\rm B} & = & \left(\frac{2c_0 a_1}{r\nu^\prime} + \frac{3h_2^{\rm B}}{r\nu^\prime} -
\frac{3k_2^{\rm B}}{2} \right)\bigg|_{r=R}\, ,  \\
\mathcal{Q}_{\rm B} & = & \frac{8M^4 K^{\rm B} - 6\mu^2}{5M}\,. 
\end{eqnarray}
Furthermore, the ellipticity can also be associated with an actual
deformation of the shape of the star as measured in terms of the
equatorial and polar radii, $R_e$ and $R_p$, and normalised to the radius
in the case of zero magnetic field, \ie
\begin{equation}
e^{\rm B} = \frac{R_e - R_p}{R}\,.
\end{equation}
Nonperturbative deformations of magnetised stars obviously require the
numerical solution of the full set of the Einstein and MHD
equations. This has has been achieved under a number of magnetic-field
configurations and strengths \cite{Bocquet1995, Cardall2001, Kiuchi2008,
  Pili2014, Chatterjee2015, Gomes2019b}.

Before moving to the next section, where we consider the perturbations
introduced by a tidal field, it is useful to summarise the results
obtained so far and make a remark. We have shown that given a perturbing
magnetic field of strength $B$, the perturbations are expressed through
the function $a_1$ that is $\mathcal{O}(B)$, so that the perturbations in
the metric, \ie $h_0^{\rm B}$, $h_2^{\rm B}$, $m_0^{\rm B}$, $m_2^{\rm
  B}$, $k_2^{\rm B}$, are all $\mathcal{O}(B^2)$. It follows from the
Einstein equations, that relate the perturbed metric with the the
perturbed energy-momentum tensor, that the magnetically perturbed energy
density and pressure $e^{\rm B}$ and $p^{{\rm B}}$ are also
$\mathcal{O}(B^2)$.

Finally, we note that the purely poloidal magnetic-field configuration
considered here has long since been shown to be unstable in generic
plasmas \cite{Tayler1973} and to lead -- over a few Alfv\'en timescales
-- to a substantial readjustment of the magnetic field in neutron
stars~\cite{Lasky2011,Ciolfi2011,Ciolfi2012}. Notwithstanding these
considerations, we employ it here because of its simplicity, which allows
for a managable analytical treatment. Furthermore, we expect that a
potentially stable configuration, e.g., as the one obtained with the
addition of a toroidal component as in a twisted-torus configuration
\cite{Ciolfi2009,Ciolfi2013}\footnote{The dynamical analysis carried out
in Refs. \cite{Ciolfi2009,Ciolfi2013} has shown that purely poloiodal (or
purely toroidal) magnetic fields are unstable. At the same time, they
have shown that once the stability has developed and has saturated, a new
mixed poloidal-toroidal magnetic-field configuration is produced. This
new configuration appears to be dynamically stable and even a small
contribution of toroidal magnetic field is sufficient to provide this
stability on dynamical timescales. This is shown, for instance, in the
middle panel of Fig. 3 of Ref. \cite{Ciolfi2013}, which reports the
evolution of poloidal and toroidal magnetic energies normalized to the
initial total magnetic energy. Note that when the instability has
stabilized after about 10 ms, the toroidal magnetic field strength is of
a few percent that of the poloidal magnetic field. Yet, this is
sufficient to avoid a new instability over the timescale of the
simulation (60 ms).}, would result in slighlty different values of the
tidal deformability, but also that the changes are not going to be more
than a factor of two for the same magnetic-field strength. Given that we
provide here a first order-of-magnitude estimate, we believe this is
reasonable compromise between a first analytical treatment and realism.

\subsection{First-order tidal-field perturbations}
\label{sec:fotf}

Next, assuming a zero magnetic field, we consider the first-order
perturbation introduced in the star by the presence of an external tidal
field, that is, we express the perturbed metric as
\begin{equation}
  \label{eq:gmunu_hTmunu}
  g_{\mu \nu} = \mathring{g}_{\mu \nu} + h^{\rm T}_{\mu \nu}\,,
\end{equation}
where the tidal-field perturbations $h^{\rm T}_{\mu \nu}$ are also
assumed to be axially symmetric (\ie with $m=0$ in a spherical-harmonic
expansion) and given by~\cite{Hinderer08,Hinderer09} 
%
\begin{eqnarray}
h_{\mu\nu}^{\rm T} & = &
\begin{pmatrix}
-e^\nu H_0 & 0 & 0 & 0 \\
0 & e^\lambda H_2 & 0 & 0 \\
0 & 0 & r^2 K & 0 \\
0 & 0 & 0 & r^2 \sin^2 \theta K \\
\end{pmatrix}P_2(\cos\theta)\,. \label{eq:pertid}
\end{eqnarray}
The resulting master equation for the tidal-field perturbations can then
be written as ~\cite{Hinderer08} (note that hereafter we will drop the
upper index ``${\rm T}$'' to allow a direct comparison with the
literature)
%
\begin{eqnarray}
& & H_0^{\prime\prime} + \left[\frac{2}{r} + \frac{2m}{r^2}e^\lambda + 4\pi r(p-e)e^\lambda \right] H_0^{\prime} + \left[4\pi e^\lambda\left(4e+8p+(p+e)\left(1+\frac{1}{c_s^2}\right)\right) -\frac{6e^\lambda}{r^2} - \nu^{\prime 2} \right] H_0  = 0\,,  \label{eq:eqtid}
\end{eqnarray}
where $c_s$ is the sound speed and the relations between $H_0(r)$ and
$H_2(r)$, $K(r)$ are given by \cite{Hinderer08}
\begin{equation}
H_2 = -H_0\,,\qquad \qquad  K^\prime = -H_0\nu^\prime - H_0^\prime\,.
\end{equation}
The behavior of the solution for $r\to 0$ is then given by
\begin{equation}
H_0(r) \to \alpha_t r^2 + \mathcal{O}(r^3)\,,
\end{equation}
while the exterior solution is
\begin{eqnarray}
  & & H_0 = c_1^e Q_2^2(z) + c_2^e P_2^2(z)\,,
  \label{eq:solt}
\end{eqnarray}
where $P_2^2$ and $Q_2^2$ are the associated Legendre functions of first
and second kind, respectively, and $c_1^e$ and $c_2^e$ are two
undetermined integration constants. By studying the behavior for $r \to
\infty$, the asymptotic behavior of the master equation is given by
\begin{eqnarray}
H_0 & = & \frac{8}{5}c_1^e\left(\frac{M}{r}\right)^3 +
\mathcal{O}\left(\left(\frac{M}{r}\right)^4\right) +
3c_2^e\left(\frac{r}{M}\right)^2
+ \mathcal{O}\left(\left(\frac{r}{M}\right)\right)\,, \label{eq:asymph0}
\end{eqnarray}
Combining now the definition of the inducing quadrupolar tidal field
$\mathcal{E}_{ij}$, with the definition of the induced quadrupole moment
$\mathcal{Q}_{ij}$, and the expansions in Eq.~\eqref{eq:moment}
\cite{Hinderer08}
\begin{equation}
-\frac{1+g_{tt}}{2} = -\frac{M}{r} - \frac{3\mathcal{Q}_{ij}}{2r^3}n^i 
n^j + \mathcal{O}\left(\frac{1}{r^4}\right) + \frac{\mathcal{E}_{ij}}{2}
r^2 n^i n^i + \mathcal{O}\left(r^3\right)\,, \label{eq:asymgtt}
\end{equation}
where $n^i := x^i/r$. The tidal deformability (or Love number) $k_2$ and
the dimensionless tidal deformability $\Lambda^{\rm T}$ can be expressed
respectively as\footnote{For this quantity only we mantain the upper
  index ${\rm T}$ so that we can reserve the symbol $\Lambda$ for the
  total dimensionless tidal deformability.} \cite{Hinderer08}
\begin{eqnarray}
  k_2 & = & -\frac{3}{2}\frac{\lambda_2}{R^5} =
  \frac{4}{15}  \frac{c_1^e}{c_2^e} \left(\frac{M}{R}\right)^5
  \,,  \label{eq:lovenum1} \\
  \Lambda^{\rm T} & := & \frac{2}{3}k_2\left(\frac{M}{R}\right)^{-5}\,. \label{eq:lovenum2}
\end{eqnarray} 

The actual numerical evaluation of these quantities takes place through
the imposition of the boundary conditions for $H_0$ and $H'_0$ at the
stellar surface, so that, in the case of a hadronic star we impose
continuity of both quantities
\begin{eqnarray}
H_0^{\rm int}(R) & = & H_0^{\rm ext}(R)\,, \label{eq:matching1n1}\\
(H_0^{\rm int})^\prime(R) & = & (H_0^{\rm ext})^\prime(R)\,, \label{eq:matching1n2}
\end{eqnarray}
while a different treatment is needed in the case of quark stars in
consideration of the discontinuity in the rest-mass density at the
stellar surface. More specifically, for quark stars we set
\cite{Damour:2009,Postnikov2010,Zhou2017}
\begin{eqnarray}
H_0^{\rm int}(R) & = & H_0^{\rm ext}(R)\,, \label{eq:matching1q1}\\
(H_0^{\rm int})^\prime(R) - \frac{4\pi R^2 e_0}{M^2}H_0^{\rm int} & = & (H_0^{\rm ext})^\prime(R)\,, \label{eq:matching1q2}
\end{eqnarray}
where $e_0$ is the energy density at the surface of the quark star. We
note that in principle we need to determine three unknowns, \ie $c_1^e,
c_2^e$, and $\alpha_t$, but have only two equations from the boundary
conditions. Fortunately, the tidal deformability depends on the ratio
$c_1^e/c_2^e$ and it is therefore possible to integrate
Eq.~\eqref{eq:eqtid} with some value of $\alpha_t$ and hence obtain --
after matching at the surface -- various pairs of values of $c_1^e$ and
$c_2^e$ for each value of $\alpha_t$; although different, they would
yield the same ratio $c_1^e/c_2^e$ and hence the same tidal deformability.

\subsection{The second-order perturbations}

Because of their linearity, the first-order perturbations introduced by
the magnetic field -- that are $\mathcal{O}(B^2)$ -- and by the tidal
field -- that are $\mathcal{O}(\mathcal{E})$ [see Eq. \eqref{eq:moment1}
  for a definition of the induced quadrupole moment $\mathcal{Q}$] -- are
decoupled and independent of each other. Hence, in order to determine how
the tidal deformability of a star is modified by the presence of a
magnetic field, it is necessary to consider higher-order perturbations
that are $\mathcal{O}(B^2 \mathcal{E})$ [see Eq. \eqref{eq:moment2} for a
  definition of the inducing quadrupole moment
  $\mathcal{E}$]. Furthermore, mathematical

In other words, at second order the perturbed metric can be expressed as
\begin{equation}
  \label{eq:gmunu_hmunu}
  g_{\mu \nu} = \mathring{g}_{\mu \nu} 
   + h^{\rm B}_{\mu \nu}
   + h^{\rm T}_{\mu \nu}
   + h^{\rm BT}_{\mu \nu}\,,
\end{equation}
where we have here implicitly neglected the second-order terms in the
magnetic field, i.e., $\mathcal{O}(B^4)$ and in the tidal field, i.e.,
$\mathcal{O}(\mathcal{E}^2)$ as these do not provide any information on
the coupling between the two effects. Mathematically, this is equivalent
to assuming that the coefficients in front of these terms are much
smaller than that of the term $\mathcal{O}(B^2 \mathcal{E})$. Once again,
we expand the metric perturbation at the second order
\eqref{eq:gmunu_hmunu} by spherical harmonic functions $Y_{\ell
  m}(\theta,\phi)$
%
\begin{eqnarray}
  \delta h_{\mu\nu} := h_{\mu\nu}^{\rm BT} =
  & = & \sum_{\ell m}
\begin{pmatrix}
-e^\nu\, \delta H_0^{\ell} & 0 & 0 & 0 \\
0 & e^\lambda \delta H_2^{\ell} & 0 & 0 \\
0 & 0 & r^2\, \delta K_{\ell} & 0 \\
0 & 0 & 0 & r^2 \sin^2\theta\, \delta K_{\ell} \\
\end{pmatrix} Y_{\ell m}(\theta,\phi)\,, \label{eq:permix}
\end{eqnarray}
%
where we have now introduced the letter ``$\delta$'' to denote any
quantity that is of second order and to avoid the use of the index
``${\rm BT}$''. The simplest case to consider at this order, which is the
one explored in this paper, consists in having the magnetic and the tidal
fields sharing the same axial symmetry, so that the axes of the magnetic
dipolar field and that of the tidal field are the same or, equivalently,
that $m=0$.

The perturbed Einstein equations with metric perturbation
\eqref{eq:permix} are given as (omitting the index ${\rm BT}$)
\begin{equation}
\delta G_{\ \mu}^{\nu} = 8\pi \delta T_{\ \mu}^{\nu}\,, \label{eq:pein}
\end{equation}
where the nonvanishing components of the perturbed energy-momentum tensor
are $\delta T_{\ 0}^{0} = -\delta p^{\rm BT}/c_s^2$ and $\delta
T_{\ i}^{i} = \delta p^{\rm BT}$, with $\delta p^{\rm BT}$ the
second-order perturbation in the pressure. The terms in the Einstein
tensor $\delta G_{\ \mu}^{\nu}$, on the other hand, can be separated into
two parts: one including terms that are the product of two first-order
perturbations (\eg $H_0 h_2^{\rm B}$), and another one which includes
purely second-order metric perturbations (\ie $\delta H_0^{\ell}$,
$\delta H_2^{\ell}$ and $\delta K_{\ell}$). Using Eqs. \eqref{eq:pein},
it is possible to find a relation between $\delta H_0^{\ell}$ and $\delta
H_2^{\ell}$ via $\delta G_{\ \theta}^{\theta} - \delta G_{\ \phi}^{\phi}
= 0$, and a similar relation can be found between $\delta
K_{\ell}^\prime$ and $\delta H_0^{\ell}$ after using $\delta
G_{\ r}^{\theta} = 0$.  Finally, using $\delta G_{\ t}^{t} - \delta
G_{\ r}^{r} = -(1/c_s^2+1)(\delta G_{\ \theta}^{\theta} + \delta
G_{\ \phi}^{\phi})/2$, and combining all the various relations, it is
possible to obtain a single master equation for $\delta
H_0^{\ell}$. After integrating out the $\theta$ dependence, and adopting
the ``polar-led'' approximation\footnote{In general, the first-order
  solutions will contribute to the second-order metric perturbations
  acting as source terms [\cf Eq. \eqref{eq:eqmix}]. In the polar-led
  approximation, the first-order contributions of the modes with
  $\ell=L\pm 2$ that impact the $\ell=L$ second-order metric
  perturbations are neglected \cite{Pani15a,Pani15b}.}, the quadrupolar
master equation for $\delta H_0$ (\ie $\delta H_0 := \delta
H_0^{\ell=2}$) can finally be written as
%
\begin{eqnarray}
& & \delta H_0^{\prime\prime} + \left[\frac{2}{r} +
    \frac{2m}{r^2}e^\lambda + 4\pi r(p-e)e^\lambda \right] \delta
  H_0^{\prime} + \left[4\pi
    e^\lambda\left(4e+8p+(p+e)\left(1+\frac{1}{c_s^2}\right)\right)
    -\frac{6e^\lambda}{r^2} - \nu^{\prime 2} \right] \delta H_0 =
  S(r)\,. \label{eq:eqmix}
\end{eqnarray}
%
Note that the terms of this master equation are arranged so that the
terms with two first-order metric perturbations (\ie perturbations we
have described in Secs. \ref{sec:fomf} and \ref{sec:fotf}) are written on
the left-hand side, while those with second-order metric perturbations on
right-hand side. Equation \eqref{eq:eqmix} is indeed very similar to
Eq.~\eqref{eq:eqtid}, with the exception of the source term $S(r)$ on the
right-hand side, which depends on the first-order solutions $H_0, K,
h_0^{\rm B}, h_2^{\rm B}, m_0^{\rm B}, m_2^{\rm B}$ and $k_2^{\rm B}$
(see Appendix \ref{sec:appendix_A} for an explicit expression).

The master equation for the exterior spacetime can be obtained easily by
requiring that there\footnote{Strictly speaking, the condition
  ${1}/{c_s^2}\rightarrow 0$ is necessary only in the case of quark
  stars, for which the energy density does not vanish at the surface. In
  this case, therefore, regularity is obtained by requiring a divergent
  sound speed.} $p,\ e,\ {1}/{c_s^2}\rightarrow 0$, and by inserting
Eqs.~\eqref{eq:solm1}--\eqref{eq:solm4} and \eqref{eq:solt} into the
source term $S$. The resulting master equation in the stellar exterior is
therefore given by
\begin{eqnarray}
& & (z^2-1)\delta H_0^{\prime\prime} + 2z\delta H_0^{\prime}
  -\left(6+\frac{4}{z^2-1} \right)\delta H_0 = S^e(r)\,,
  \label{eq:eqmixr}
\end{eqnarray}
where $S^e(r)$ is obviously the source term in the stellar exterior.

Equation \eqref{eq:eqmixr} can not be solved analytically and so
numerical methods have to be employed to analyze its asymptotic behavior
for $r \to +\infty$. In analogy with Eq. \eqref{eq:solt}, we can express
the general solution of Eq.~\eqref{eq:eqmixr} as
\begin{eqnarray}
  & & \delta H_0(z) = d_1^e Q_2^2(z) + d_2^e P_2^2(z) + \delta \hat{H}(z)\,,
  \label{eq:solmix}
\end{eqnarray}
where $d_1^e$ and $d_2^e$ are free constants to be determined, and
$\delta \hat{H}(z)$ is a special solution of this differential equation
that can be obtained numerically with arbitrary initial
condition. Because the asymptotic behavior of the solution is unknown
when $\delta \hat{H}(z)$ is solved merely numerically, we can first
analyze the behaviour of the general function \eqref{eq:solmix} for large
$r$. In this case, the exterior source term $S^e$ can be split into two
terms, \ie $S^e = c_1^e S_1 + c_2^e S_2$, where, for $r \rightarrow
+\infty$ we have (see Appendix \ref{sec:appendix_A} for the expression of
$S^e$)
\begin{equation}
  S_1 \rightarrow -\frac{144 c_2}{5M} \left(\frac{M}{r} \right)^4\,, \qquad
  S_2 \rightarrow \frac{8c_2}{M} \left(\frac{r}{M}\right)\,.
\end{equation}
The special solution at large $r$ can then be written as
\begin{eqnarray}
\delta \hat{H}(z) & = & c_1^e \frac{6c_2}{5M}\left(\frac{M}{r} \right)^3
+ \mathcal{O}\left(\left(\frac{M}{r} \right)^4 \right)
+ c_2^e \frac{2c_2}{3M} \left(3\frac{r}{M}-1 \right) + \mathcal{O}\left(\frac{M}{r} \right)\,. \label{eq:asymdh}
\end{eqnarray}
In practice, we solve numerically Eq.~\eqref{eq:eqmixr} twice, having as
source term either $S^e = S_1$ or $S^e = S_2$. In doing so, we take
expressions~\eqref{eq:asymdh} as initial conditions to integrate the
differential equation \eqref{eq:eqmixr} from infinity to the stellar
surface, obtaining as final general solution the expression
\begin{eqnarray}
& & \delta H_0(z) = d_1^e Q_2^2(z) + d_2^e P_2^2(z) + c_1^e \delta
  \hat{H}_1(z) + c_2^e \delta \hat{H}_2(z)\,, 
  \label{eq:solmixr}
\end{eqnarray}
where $\delta \hat{H}_1(z)$ and $\delta \hat{H}_2(z)$ are the numerical
solutions for $S^e=S_1$ and $S^e=S_2$, respectively. Next, from the
asymptotic behavior of the $tt$ component of metric \eqref{eq:asymgtt},
we can calculate the inducing quadrupolar tidal field $\mathcal{E}$ and
the corresponding induced quadrupole moment $\mathcal{Q}$ after 
collecting all tidal metric perturbation terms \eqref{eq:asymph0} and 
\eqref{eq:asymdh}, and writing down $\mathcal{E}$ and $\mathcal{Q}$ as
\begin{eqnarray}
& & \mathcal{E} := \mathcal{E}_{ij}n^i n^j = \frac{6}{M^2} c_2^e + \frac{6}{M^2} d_2^e\,, \label{eq:moment1} \\
& & \mathcal{Q} := \mathcal{Q}_{ij}n^i n^j = -\frac{16M^3}{15}(c_1^e + d_1^e) - \frac{4M^3}{5}\frac{c_2}{M} c_1^e. \label{eq:moment2}
\end{eqnarray}
Since the quadrupolar tidal field $\mathcal{E}$ is sourced from an
exterior tidal field (\ie that produced by the companion star), it should
not be affected by the order at which the interior solution is
computed. To reflect this behaviour, the integration constant $d_2^e$
should vanish. Finally, the second-order magnetically modified
even-parity tidal quadrupolar deformability (or simply ``magnetic tidal
deformability'') can be written as
\begin{eqnarray}
  & & \delta k_2 := \frac{1}{5}\left(\frac{M}{R}\right)^5\left(
  \frac{4}{3} \frac{d_1^e}{c_2^e} +
  \frac{c_2}{M}\frac{c_1^e}{c_2^e} \right)\,, \label{eq:dk2} \\
&&  \delta \Lambda :=  \frac{2}{3} \delta k_2\left(\frac{M}{R}\right)^{-5}\,. \label{eq:dlovenum2}
\end{eqnarray}

The ratio of the two constants $d_1^e/c_2^e$ is determined by matching
the interior solution [Eq.~\eqref{eq:eqmix}] with the exterior one
[Eq.~\eqref{eq:solmixr}] via the continuity of $\delta H_0$ and $\delta
H^{\prime}_0$ across the stellar surface, \ie
\begin{eqnarray}
\delta H_0^{\rm ext}(R) & = & \delta H_0^{\rm int}(R)\,, \label{eq:matching2q1} \\
(\delta H_0^{\rm ext})^\prime(R) & = & (\delta H_0^{\rm int})^\prime(R) - \frac{4\pi R^2 e_0}{M^2}\delta H_0^{\rm int}(R) + S_{\rm surf}\,, \label{eq:matching2q2}
\end{eqnarray}
where $S_{\rm surf}$ is the contribution from the source term at the
stellar surface and will be shown explicitly in Appendix
\ref{sec:appendix_A}. Note that the second and third terms on the
right-hand side of \eqref{eq:matching2q2} are needed only in the case of
a quark star and are zero for a standard hadronic star. Note that since
expressions \eqref{eq:dk2} and \eqref{eq:dlovenum2} represent the
second-order corrections only, the total tidal deformability for a
magnetised neutron star is given by
\begin{eqnarray}
  \label{eq:total}
&&  \lambda_2 := - \frac{\mathcal{Q}_{ij}}{\mathcal{E}_{ij}} =
  -\frac{2}{3} R^5 (k_2 + \delta k_2)\,, \\
&&  \Lambda = \Lambda^{\rm T} + \delta \Lambda :=  \frac{2}{3} (k_2 + \delta k_2) \left(\frac{M}{R}\right)^{-5} \,. 
\end{eqnarray}

A few remarks before moving to the next section. First, while $k_2$ and
$\delta k_2$ both measure the quadrupolar even-parity tidal deformability
of a star in the external tidal field of a companion, they depend on
different quantities. More specifically, while $k_2=k_2(M,R)$, where $M$
and $R$ are the stellar mass and radius, $\delta k_2= \delta k_2(M,R,B)$,
so that $\delta k_2 \to 0$ for $B \to 0$. Second, as we will see in the
following, $\delta k_2 \ll k_2$ unless extremely strong magnetic fields
are considered. Finally, while $k_2$ is always positive, $\delta k_2$ can
change sign, although $\lambda_2$ will remain positive.

\section{Numerical results and physical implications}
\label{sec:results}

In what follows we discuss the results of the numerical solution of the
perturbative equations discussed in the previous sections, paying
attention to the magnitude of the magnetic tidal deformability
(\ref{sec:IIIA}), on its impact on the GW-phase evolution in binary
systems (\ref{sec:IIIB}), on how it compares with spin-induced
corrections (\ref{sec:IIIC}), and, finally, under what conditions the
I-Love relations break-down (\ref{sec:IIID}).

\subsection{Tidal deformability for magnetised neutron and quark stars}
\label{sec:IIIA}

We have already discussed briefly in the previous sections about the
numerical solution of the perturbative equations. In essence, we first
solve simultaneously the TOV equations \eqref{eq:tov1}--\eqref{eq:tov3}
and the first-order perturbative equations
\eqref{eq:eqm11}--\eqref{eq:eqm32}, \eqref{eq:eqtid}. Making use of the
computed zeroth- and first-order solutions, the second-order master
equation \eqref{eq:eqmix} is solved with the initial condition $\delta
H_0(r\simeq 0) = r^2 + \mathcal{O}(r^3)$. The solution obtained
numerically in this way is denoted by $\delta H_0^{\rm N}$, and the
general solution of Eq.~\eqref{eq:eqmix} can be written in the form of
\begin{eqnarray}
\delta \tilde{H}_0 & := & c^{\rm BT} \delta H_0^{S=0} + \delta H_0^{\rm N} \,, \label{eq:gensol}
\end{eqnarray}
where $\delta H_0^{S=0}$ is the solution of Eq.~\eqref{eq:eqtid} [or,
  equivalently, of Eq.~\eqref{eq:eqmix} with vanishing source term
  $S(r)$], and $c^{\rm BT}$ is a constant that is determined, together
with $d_1^e$, via the boundary conditions at the stellar surface [\cf
  Eqs.~\eqref{eq:matching2q1}--\eqref{eq:matching2q2}].

For the zeroth-order solutions we consider eight different EOSs that
serve to illustrate the behaviour across different tidal
deformabilities. In particular, we compute equilibrium models for neutron
stars described by the EOSs: ${\rm WFF1}$ \cite{Wiringa88}, ${\rm APR}$
\cite{Akmal1998a}, ${\rm SLy4}$ \cite{Gulminelli2015}, ${\rm qmf18}$
\cite{Zhu2018} and ${\rm MPA1}$ \cite{Muether87}. All of these EOSs can
fulfil the constraints of a maximum mass above two solar masses
\cite{Demorest2010,Antoniadis_fulllist:2013} and have tidal
deformabilities in broad agreement with the constraints and their
uncertainties derived from GW170817
\cite{Abbott2017_etal,Abbott2018b}. In addition, we also consider two
EOSs describing quark stars, namely, CIDDM \cite{Qauli2016} and MIT2cfl
\cite{Zhou2017}, where the latter is obtained through the MIT bag model
with parameters $\Delta = 100\ {\rm MeV},\ B_{\rm eff}^{1/4} = 150\ {\rm
  MeV},\ m_s = 100\ {\rm MeV}$, and $a_4 = 0.61$ (see \cite{Zhou2017} for
more details). Also these quark-star EOSs satisfy the constraint of
having maximum masses above two solar masses.

\begin{figure*}[t]
  \centering
  \includegraphics[width=0.49\textwidth]{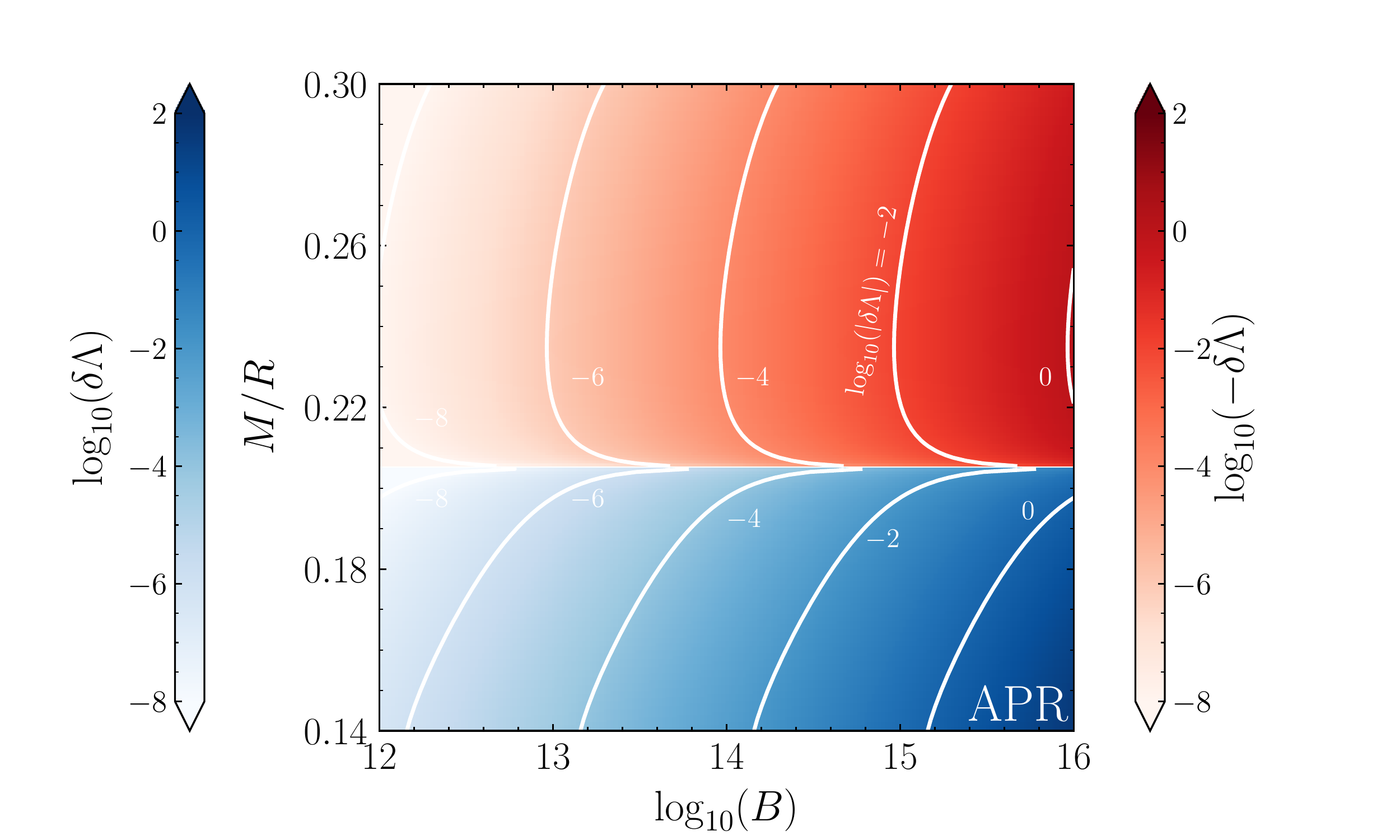}
  \includegraphics[width=0.49\textwidth]{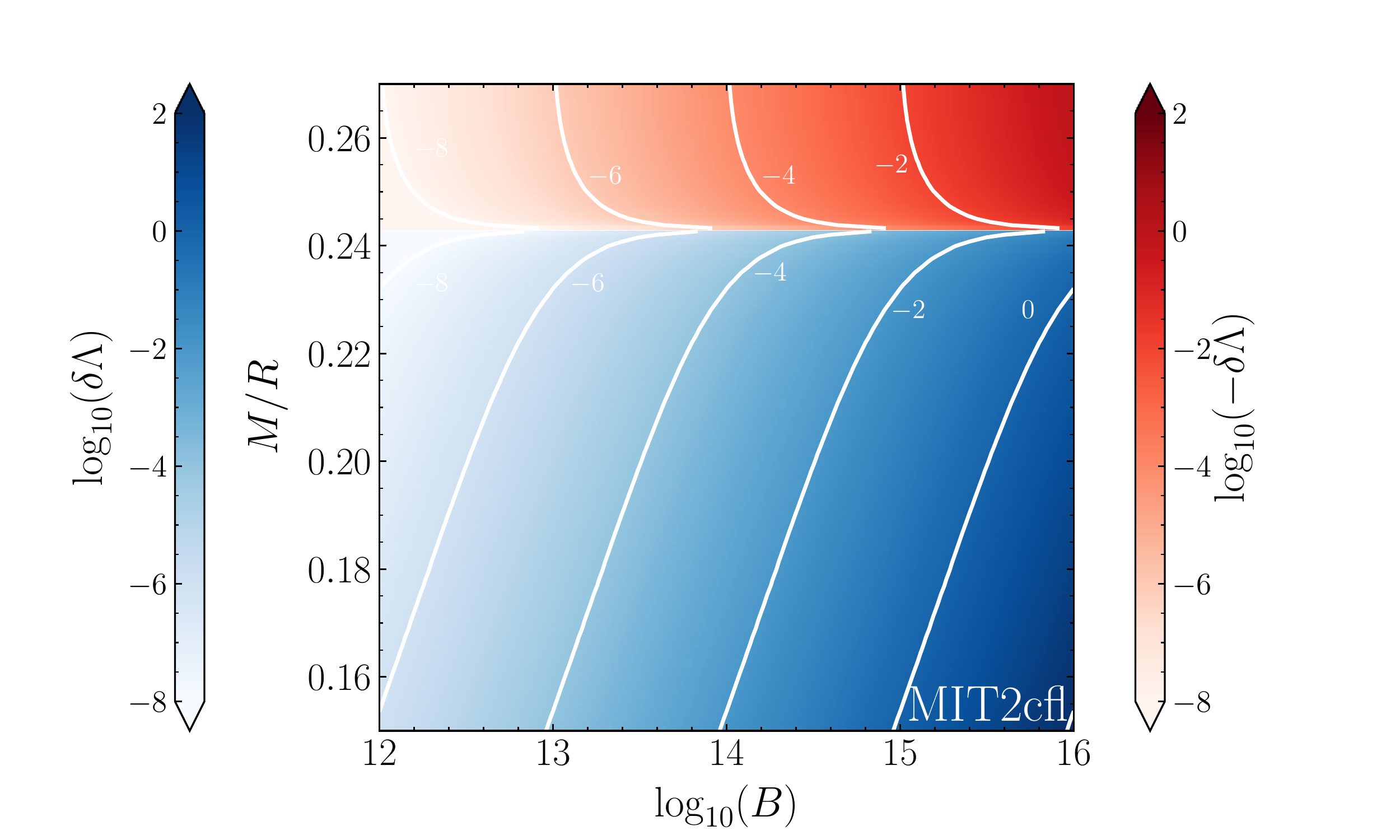}
  \caption{Dimensionless magnetically modified tidal deformability
    $\delta \Lambda$ shown as a function of magnetic field strength and
    compactness $M/R$ of the star. The left panel refers to the
    neutron-star EOS APR, while the right one to the EOS MIT2cfl, and is
    therefore representative of quark stars. Two different colours are
    used account for the different signs of $\delta \Lambda$ and indicate
    that while a magnetic field increases the tidal deformation for weak
    gravitational fields, it opposes it in stronger
    gravity.}\label{fig:dk2}
\end{figure*}

The results of the numerical integration of the magnetically modified
dimensionless tidal deformability (or simply ``dimensionless magnetic
tidal deformability'') $\delta \Lambda$ are shown in Fig. \ref{fig:dk2}
as a function of magnetic-field strength at the stellar pole for neutron
stars with the APR EOS (left panel) and for quark stars with the MIT2cfl
EOS (right panel). Note that because the dimensionless magnetic tidal
deformability can change sign for sufficiently large compactnesses, we
report, respectively in blue and red, the positive and negative values of
$\delta \Lambda$. Note also that the sign change takes place at
essentially a constant value of the stellar compactness, \ie at $M/R
\simeq 0.205$. The existence of such a zero can be easily deduced from
the functional form of $\delta k_2$ as given in Eq. \eqref{eq:dk2}: since
the constant $c_2$ is proportional to the stellar radius and because the
ratio $d^e_1/c^e_2$ become negative above a certain compactness,
expression \eqref{eq:dk2} highlights that the magnetic tidal
deformability will be zero for a given compactness. From a more physical
point of view, the behaviour shown in Fig. \ref{fig:dk2} highlights the
fact that for weak gravitational fields (\ie for small $M/R$), the
presence of a magnetic field simply enhances the tidal deformability as
the quadrupolar deformation introduced by the magnetic field adds
positively to that introduced by the tidal field. However, for strong
gravitational fields (\ie for large $M/R$), the opposite is true and the
magnetic field prevents -- via the additional magnetic pressure and
magnetic tension -- a quadrupolar deformation.

This behaviour can also be found in quark stars (right panel of
Fig. \ref{fig:dk2}), although the change in sign in $\delta \Lambda$
takes place at much larger masses and compactnesses (\ie $M \simeq
2.03\ M_\odot,\ M/R \simeq 0.245$ for the MIT2cfl EOS). Furthermore, in
quark stars, $\delta \Lambda$ decreases monotonically with increasing
compactness. These different behaviours at low compactnesses is most
likely due to the different behaviour of the outer layers of the two
stellar types. In general, in fact, the crust of neutron star follows an
EOS that is very different from that of the core. On the other hand, by
lacking a crust, quark stars have a behaviour that does not change with
compactness and hence yields a magnetic tidal deformability that is
mostly positive.

Note also that since $a_1,\mu \propto B$ [\cf Eqs. \eqref{eq:mag1} and
  \eqref{eq:a1}], it follows that $m_0^{\rm B}, P_0 \propto B^2$ [\cf
  Eqs.  \eqref{eq:eqm31}--\eqref{eq:eqm32}], so both the constants $c_2$
and $d_1^e$ are proportional to $B^2$ [\cf Eqs. \eqref{eq:eqmix} and
  \eqref{eq:solmix}]. As a result, the behaviour of $\delta \Lambda$ as a
function of the magnetic field reported in Fig. \ref{fig:dk2} is actually
a linear one. Overall, for the APR EOS, the maximum value of the magnetic
tidal deformability is $\delta \Lambda = 53.9$ and is reached at $M/R =
0.133$ for a magnetic field of $B=10^{16}\,{\rm G}$; this is roughly
$4\%$ of $\Lambda^{\rm T}$; on the other hand, for the MIT2cfl the value
is $\delta \Lambda = 288.7$ at $M/R = 0.133$ for a magnetic field of
$B=10^{16}\,{\rm G}$; this is roughly $6\%$ of $\Lambda^{\rm T}$.

\begin{figure*}[h!]
  \centering
  \includegraphics[width=0.49\textwidth]{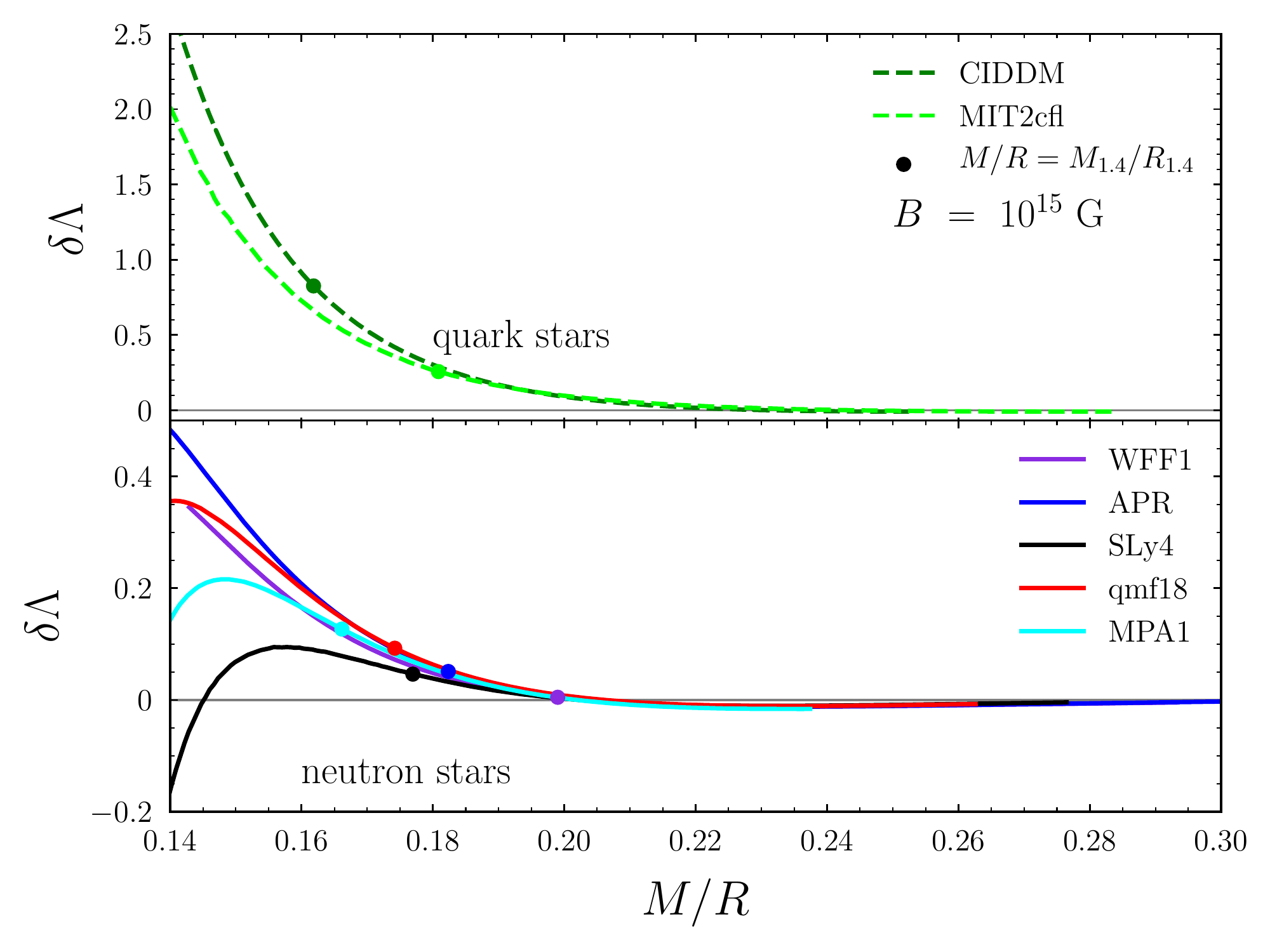}
  \includegraphics[width=0.49\textwidth]{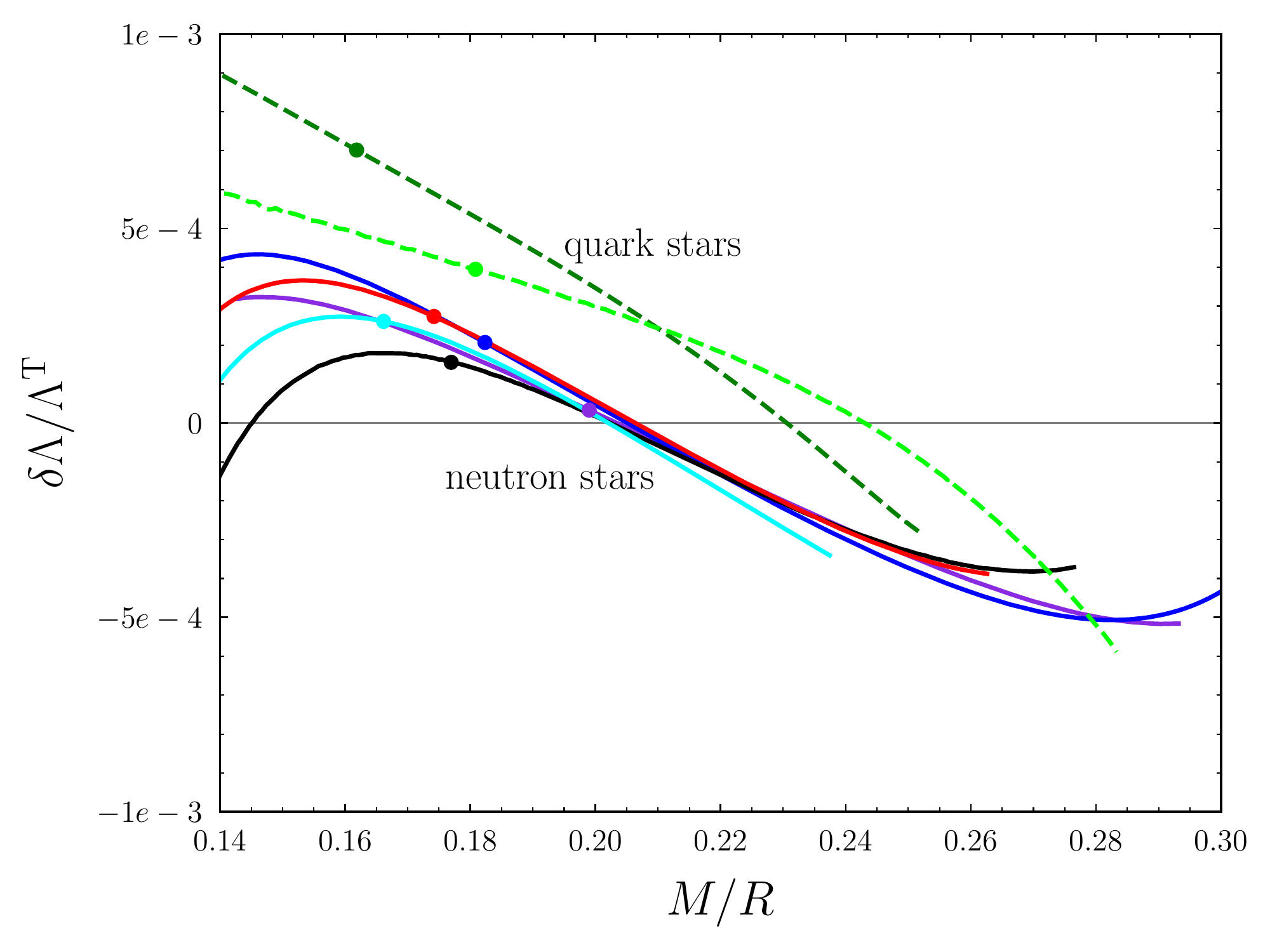}
  \caption{\textit{Left panel:} Dimensionless magnetic tidal
    deformability $\delta \Lambda$ shown as a function of the stellar
    compactness and for a fixed magnetic field of $B=10^{15}\,{\rm
      G}$. The top part refers to quark-star EOSs, while the bottom one
    to representative neutron-star EOSs. Marked with dots are the
    positions of stars with $M=1.4\ M_\odot$. \textit{Right panel:}
    Relative weight of the dimensionless magnetic tidal deformability
    $\delta \Lambda$ when compared with the tidal deformability
    $\Lambda^{\rm T}$. Different lines refer to different EOSs.}
  \label{fig:eos}
\end{figure*}

Figure \ref{fig:eos} provides a different view of the dependence of
magnetic tidal deformability by reporting in the left panel $\delta
\Lambda$ as a function of the stellar compactness for various EOSs
relative to neutron stars (bottom part) and quark stars (top part), with
a filled circle marking the reference value of the compactness of a star
with $M=1.4\,M_{\odot}$. The data in the figure refers to a reference
magnetic field of $B=10^{15}\,{\rm G}$ but, obviously, larger/smaller
values would be obtained for $\delta \Lambda$ when considering
larger/smaller values of $B$. Note the very different behaviour between
the two types of stars, with $\delta \Lambda$ having a local maximum in
the case of neutron stars, while decreasing monotonically for increasing
compactness in the case of quark stars. More importantly, note that the
modification of the tidal deformability for quark stars is significantly
larger, being even 20 times larger than that of neutron stars. Overall,
the different magnitude and dependence on the stellar compactness could
provide an important signature to distinguish between the two classes of
stars.

Shown instead in the right panel of Fig. \ref{fig:eos} is the relative
change of the tidal deformability, $\delta \Lambda/\Lambda^{\rm T}$,
highlighting that the magnetically induced corrections to the tidal
deformability are normally only a small fraction of the ordinary tidal
deformation, \ie $\lesssim 10^{-3}$ for magnetic field as large as $\sim
10^{15}\,{\rm G}$ and $\lesssim 10^{-9}$ for more realistic magnetic
fields of $\sim 10^{12}\,{\rm G}$.

\subsection{Impact of the phase evolution in binary systems}
\label{sec:IIIB}

In order to study the impact that the magnetic tidal deformability has on
the evolution of the GW signal from merging binaries, we have computed
the GW-phase evolution of representative binaries for the various EOSs
considered here and contrasted the situations in which the magnetic field
is either zero or not. We recall that GW waveforms of inspiralling
binaries are normally calibrated by fitting the numerical-relativity
results of the late-inspiral and merger phases, so they can extend the
waveforms essentially up to merger (see
\cite{Hinderer2018b,Dietrich2020b} for two recent reviews). Generally,
the most common semi-analytical models are the phenomenological
(``Phenom'') models -- which combine in a phenomenological manner and at
different frequencies the PN evolution with the one from numerical
simulations \cite{Ajith:2007kx:longal, Hannam2013b, Khan2016} -- and the
effective-one-body (``EOB'') models -- which convert the binary inspiral
two-body problem to a one-body problem of describing a test particle
moving in a deformed black-hole spacetime \cite{Buonanno:1998gg}. Two
different and independent EOB models are being developed in the
literature, \ie the SEOBNRv4 \cite{Bohe2017, Cotesta2018} and the
TEOBResumS models, \cite{Nagar2018,Nagar2019}, and their differences are
discussed in Ref. \cite{Piero2019}. There are two different ways that the
tidal contribution to the waveform are take into account: It can be
incorporated directly into EOB formalism in the case of TEOBResumS and
SEOBNRv4 models. Alternatively, it can also appear as an additional
correction to the tide-free expression for the GW-phase evolution in the
case of the SEOBNRv4 and Phenom models.

For convenience, we have here employed the tidal model NRTidal
\cite{Dietrich2019}, to calculate the contribution of tidal deformability
to the GW-phase evolution, while the IMRPhenomD model \cite{Khan2016} is
used to handle the black-hole binary part of the inspiral. In practice,
we have employed the publicly available \texttt{PyCBC} software
\cite{alex2020} to generate the waveforms produced by an equal-mass
binary of compact stars with single mass $M=1.4\ M_{\odot}$, magnetic
fields of various strength, starting from an initial frequency of
$60\,{\rm Hz}$ and up to the merger time. In this way, it is possible to
define the GW phase differences between the tidal effects with and
without magnetic field as
\begin{equation}
\Delta \phi(t) :=  \left.\phi(t)\right|_{B\neq0} - \left.\phi(t)\right|_{B=0}\,.
\end{equation}

\begin{figure*}[!t]
  \centering
  \includegraphics[width=0.49\textwidth]{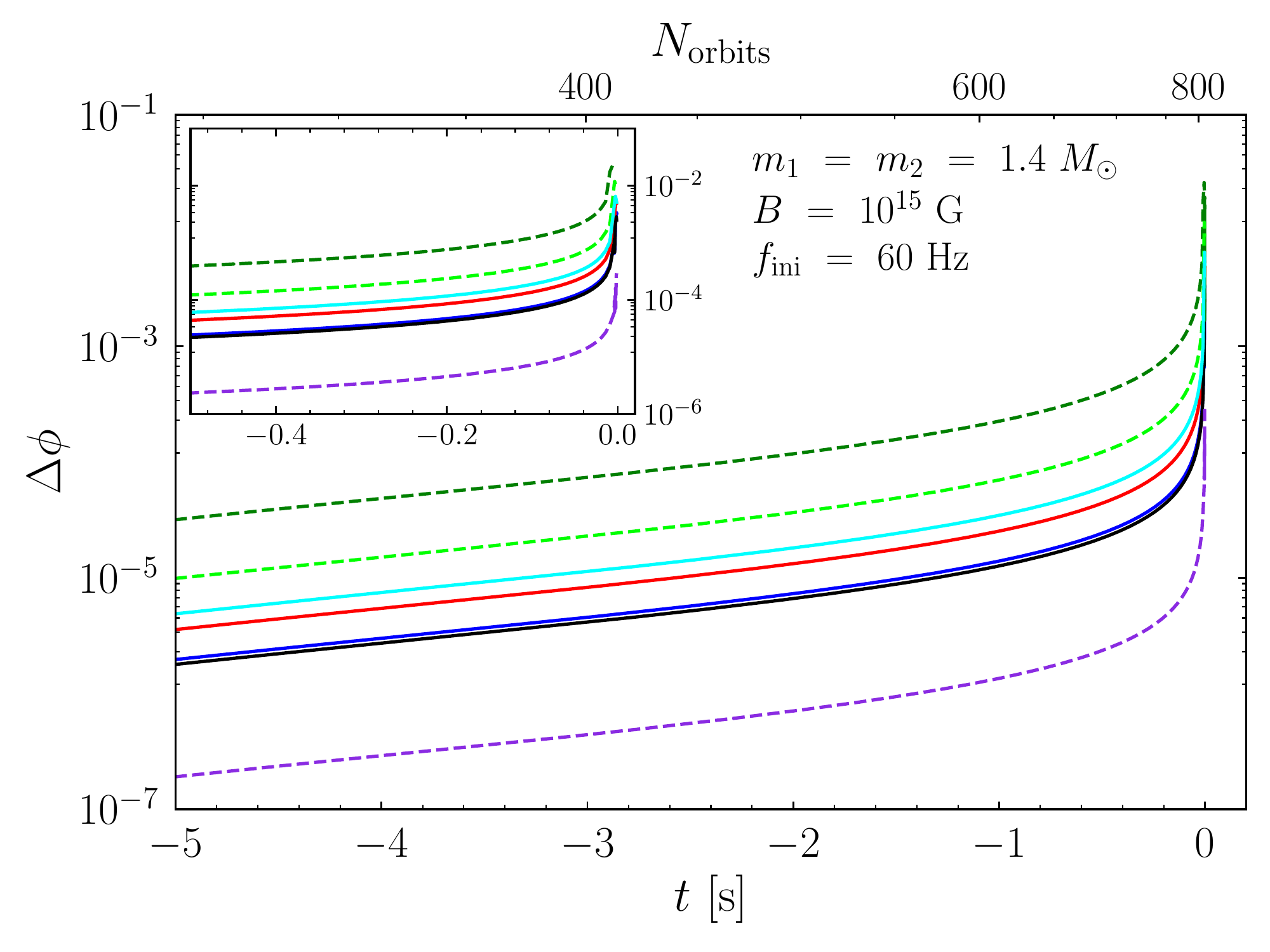}
  \includegraphics[width=0.49\textwidth]{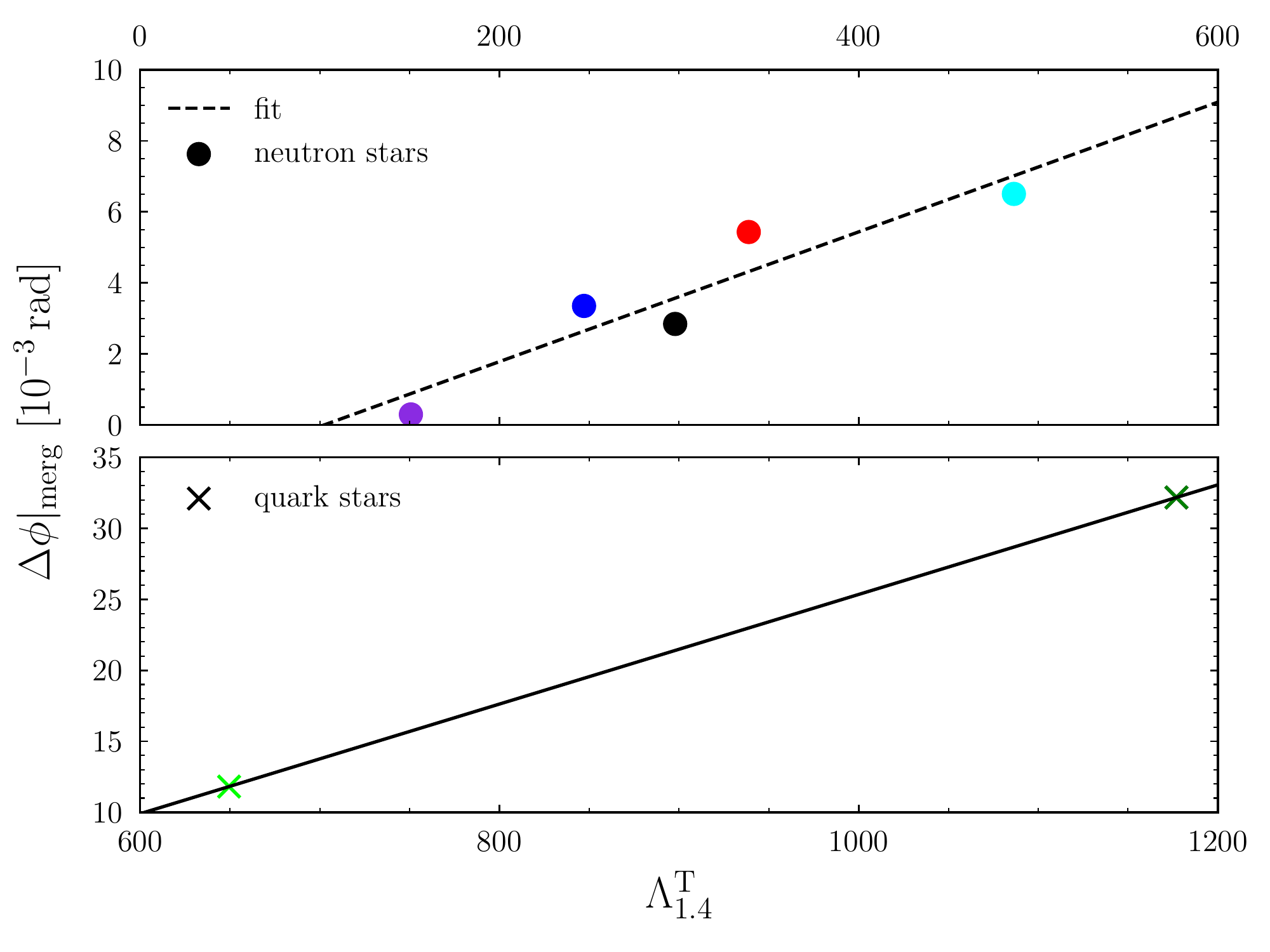}
  \caption{\textit{Left panel:} Evolution of the GW phase differences
    $\Delta \phi$ relative to binaries with zero magnetic field or with
    $B=10^{15}\,{\rm G}$. Different lines refer to different EOSs, but
    are all relative to a binary with masses $m_1=m_2=1.4\,M_{\odot}$,
    that enters the detector at an initial frequency of $60\,{\rm
      Hz}$. The inset shows a magnification near the
    merger. \textit{Right panel:} Final phase difference at the merger
    $\Delta \phi\vert_{\rm merg}$ shown as a function of the tidal
    deformability of a star with $M=1.4\,M_{\odot}$. The top part refers
    neutron stars, while the bottom part to quark stars; the colour code
    used is the same as in the left panel and hints to a linear behavior
    for neutron stars.}
    \label{fig:phase}
\end{figure*}

Figure \ref{fig:phase} reports in its left panel the evolution of the
phase difference for a reference magnetic field $B=10^{15}\, {\rm G}$ and
for different EOSs relative to either neutron stars (solid lines) or
quark stars (dashed lines), using the same colour convention as in
Fig. \ref{fig:eos}. Note that the phase differences are computed up to
the merger frequency, which was shown to follow a universal relation with
the tidal deformability $\Lambda^{\rm T}$
\cite{Read2013,Bernuzzi2015a,Takami2015,Rezzolla2016} in the case of
hadronic stars. It is presently unclear if such universal relations hold
also for quark stars and, if so, whether they have the same functional
behaviour. Since the \texttt{PyCBC} software does not discriminate
between the two classes of compact stars, we have used the same universal
relations to compute the GW signal of quark stars up to the presumed
merger frequency.

Not surprisingly, the growth of the phase difference reported in the left
panel of Fig. \ref{fig:phase} is very small apart from the final
fractions of a second preceding the merger (see inset). This is obviously
due to the fact that tidal effects become important only when the two
compact stars have reached a very small separation. Note also that
magnetised quark stars yield much large dephasing, which can be one or
even two orders of magnitude larger than the corresponding one obtained
in the case of neutron stars. Also in this case, however, such changes
are comparatively large because of the large reference magnetic fields,
so that the values reported serve mostly as upper limits.

Shown instead in the right panel of Fig. \ref{fig:phase} is the final
GW-phase difference at merger for a reference magnetic field of
$10^{15}\,{\rm G}$, different EOSs, and when shown as a function of the
tidal deformability of a $1.4\,M_{\odot}$ star $\Lambda^{\rm
  T}_{1.4}$. The upper part of the panel refers to neutron stars (filled
circles), while the lower part to quark stars (crosses). Furthermore,
while $\Delta \phi_{\rm merg} = \mathcal{O}\left(10^{-3}\right)\,{\rm
  rad}$ for such a large magnetic field, much smaller phase differences
are measured for more realistic magnetic fields, with an overall trend
$\Delta \phi_{\rm merg} \sim B^{2}$. As a result, exploiting the overall
behaviour shown by the neutron-star EOSs considered here, it is possible
to recognise a linear dependence of the maximum phase difference of the
type $\Delta \phi_{\rm merg} = a + b \,\Lambda^{\rm T}_{1.4}$, with
$a=-1.873\, (B/10^{15}\,{\rm G})^2$ and $b=0.018 \, (B/10^{15}\,{\rm
  G})^2$.

\begin{figure*}[!t]
  \centering
  \includegraphics[width=0.49\textwidth]{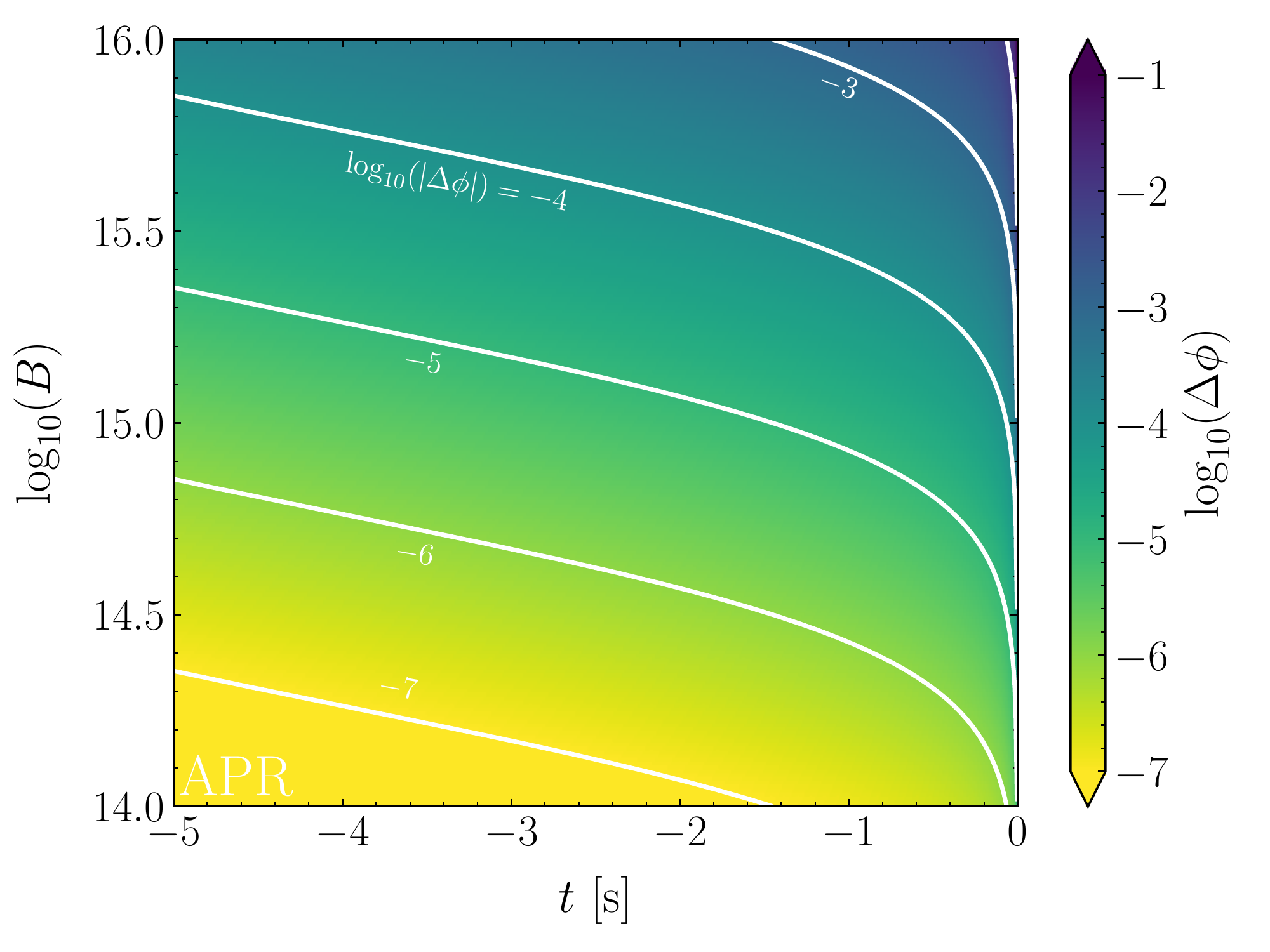}
  \includegraphics[width=0.49\textwidth]{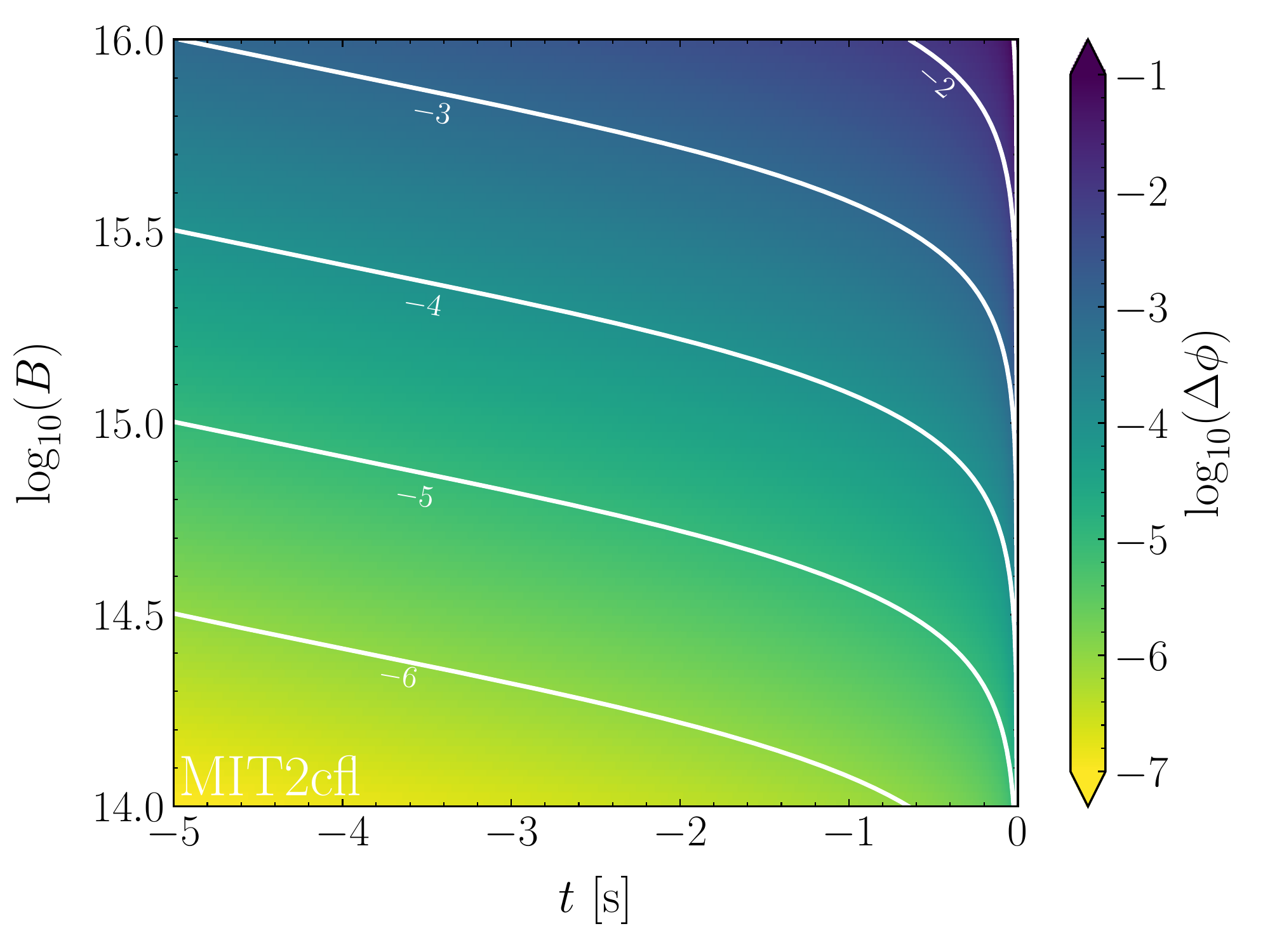}
  \caption{Evolution of the GW phase differences $\Delta \phi$ shown as a
    function of different magnetic-field strengths. Also in this case,
    the data refers to a representative binary with masses
    $m_1=m_2=1.4\,M_{\odot}$, that enters the detector at an initial
    frequency of $60\,{\rm Hz}$. The left panel refers to the
    neutron-star EOS APR, while the right one to the EOS MIT2cfl, and is
    therefore representative of quark stars. }
\label{fig:btph}
\end{figure*}

Figure \ref{fig:btph} shows the phase difference as function of time and
of the magnetic-field strength in the range from $10^{14}\, {\rm G}$ to
$10^{16}\,{\rm G}$. The left and right panels refer to the APR and to the
MIT2cfl EOSs, respectively. Also in this case, we stress that these
magnetic fields are considered here not because they are particularly
realistic, but because they serve to set stringent upper limits on the
impact that magnetic fields may have on the GW-phase evolution. In
particular, assuming the extreme case of a magnetic field $B=
10^{16}\,{\rm G}$, the final phase difference at merger is $\Delta
\phi_{\rm merg} \lesssim 0.65\,{\rm rad}$ for the neutron-star EOSs
considered here, and $\Delta \phi_{\rm merg} \lesssim 3.2\,{\rm rad}$ for
quark-star EOSs. All of this information, together with the
representative values of the magnetic tidal deformabilities, are
summarised in Table \ref{tab:ph}.

\begin{table}[t]
  \renewcommand{\arraystretch}{1.3}
  \caption{Summary of the most important quantities computed here for the
    various EOSs considered. Reported in the various columns are: the
    dimensionless tidal deformability $\Lambda^{\rm T}$, the
    dimensionless magnetic deformability $\delta \Lambda$, its relative
    weight with respect to the dimensionless tidal deformability $\delta
    \Lambda/\Lambda^{\rm T}$, the final phase differences at merger
    $\Delta \phi_{\rm merg}$. All quantities are computed for a reference
    magnetic field of $B=10^{15}\,{\rm G}$ and a stellar mass of
    $1.4\,M_\odot$, whose corresponding stellar radius $R_{1.4}$ is
    reported on the last column.}
\begin{center}
\begin{tabular}{|l|r|r|r|r|r|} 
  \hline
  \hline
  & $\Lambda^{\rm T}_{1.4}$ & $\left(\delta \Lambda\right)_{1.4}$ & 
  $\left(\delta \Lambda/\Lambda^{\rm T}\right)_{1.4}$ & $\Delta \phi\vert_{\rm merg}$ & $R_{1.4}$  \\ 
            &   & $[10^{-2}]\ \ $  & $[10^{-4}]\ \ \ \ \ $ & $[10^{-3}\,{\rm rad}]$ & $[{\rm km}]\ $    \\ 
  \hline                                                                                   
${\rm WFF1}$    & $151$   & $0.495$   &  $0.328$   & $0.294$                &  $10.39$      \\
${\rm APR}$     & $247$   & $5.109$   &  $2.067$   & $3.350$                &  $11.34$      \\
${\rm SLy4}$    & $298$   & $4.636$   &  $1.557$   & $2.843$                &  $11.68$      \\
${\rm qmf18}$   & $339$   & $9.273$   &  $2.737$   & $5.432$                &  $11.87$      \\
${\rm MPA1}$    & $486$   & $12.693$  &  $2.609$   & $6.505$                &  $12.44$      \\
${\rm CIDDM}$   & $1177$  & $82.557$  &  $7.014$   & $32.168$               &  $12.77$      \\
${\rm MIT2cfl}$ & $650$   & $25.653$  &  $3.950$   & $11.813$               &  $11.43$      \\
  \hline
  \hline
\end{tabular}
\label{tab:ph}
\end{center}
\end{table}

\subsection{Comparison with other high-order corrections}
\label{sec:IIIC}

In addition to the corrections introduced by the presence of a magnetic
field, there are also some other high-order corrections to the tidal
deformability that can have an impact on the GW emission.  In particular,
as anticipated in Sec. \ref{sec:introduction}, given an odd-parity
external quadrupolar tidal field, $\mathcal{S}_{ij}$, there will be an
odd-parity response of the star in terms of the stellar mass-current
quadrupole moment.  This tidal deformability can be obtained by looking
at the $g_{tj}$ component of metric at a large distance $r$ from the star
\cite{Hinderer2018b}
\begin{eqnarray}
g_{tj} & = &  -\frac{8}{r^3}\epsilon_{ijp}\mathcal{S}^p_{\ k} n^{<ki>} + \mathcal{O}\left(r^{-4}\right)
+ \frac{2}{3}\epsilon_{jpq}\mathcal{B}^q_{\ k}\, r^2 n^{<pk>} + \mathcal{O}\left(r^{3}\right)\,, \label{eq:magmet}
\end{eqnarray}
where $\mathcal{S}_{ij}$ is the stellar mass-current quadrupole moment,
$\mathcal{B}_{ij}$ is the odd-parity induced quadrupolar tidal field, and
$n^{<ki>}:= n^k n^i - \delta^{ki}$ is the symmetric and trace-free
projection tensor.

As discussed in Sec. \ref{sec:introduction}, the odd-parity quadrupolar
tidal deformability $\sigma_2$ can be defined as the ratio [\cf
  Eq. \eqref{eq:mass-current}]
\begin{eqnarray}
 \sigma_2 := \frac{\mathcal{S}_{ij}}{\mathcal{B}_{ij}}\,,  \label{eq:magmoment}
\end{eqnarray}
from which it is possible to build a dimensionless odd-parity tidal
deformability $j_2$
\begin{eqnarray}
  j_2 := \frac{48}{R^5}\sigma_2 = \frac{48}{R^5}\frac{\mathcal{S}_{ij}}{\mathcal{B}_{ij}}
  \,, \label{eq:j2}
\end{eqnarray}
which appears as a correction to the GW-phase evolution at 6PN order, in
contrast with the even-parity tidal deformability, which appears at 5PN
order (see Refs. \cite{,Yagi2014c,Batoul2018,JimenezForteza2018} for more
details).

We note that even in the absence of a magnetic field, other high-order
corrections to the GW-phase evolution emerge if the tidally deformed star
is rotating. In this case, in fact, couplings appear between multipole
moments of different parity. For instance, the odd-parity octupole tidal
field could produce an even-parity mass quadrupole moment, and the
even-parity octupole tidal field could induce an odd-parity mass-current
quadrupole moment \cite{Pani15a,Pani15b}. The corresponding tidal
deformabilities are denoted as $\lambda_{23}$ and $\sigma_{23}$ and
contribute to the GW-phase evolution starting from the 6.5PN order, which
is also the order at which the corrections from the coupling of the
stellar spin with the even-parity tidal deformability also
emerge. However, because the inclusion of these rotational corrections in
the Lagrangian formulation of the binary dynamics remains conceptually
unclear \cite{Abdelsalhin2018a, JimenezForteza2018}, they will be ignored
here, as done in \cite{JimenezForteza2018} (see also \cite{Poisson2020b}
for an additional source of concern on the magnitude of these
corrections).

\begin{figure*}[!t]
  \centering
  \includegraphics[width=0.49\textwidth]{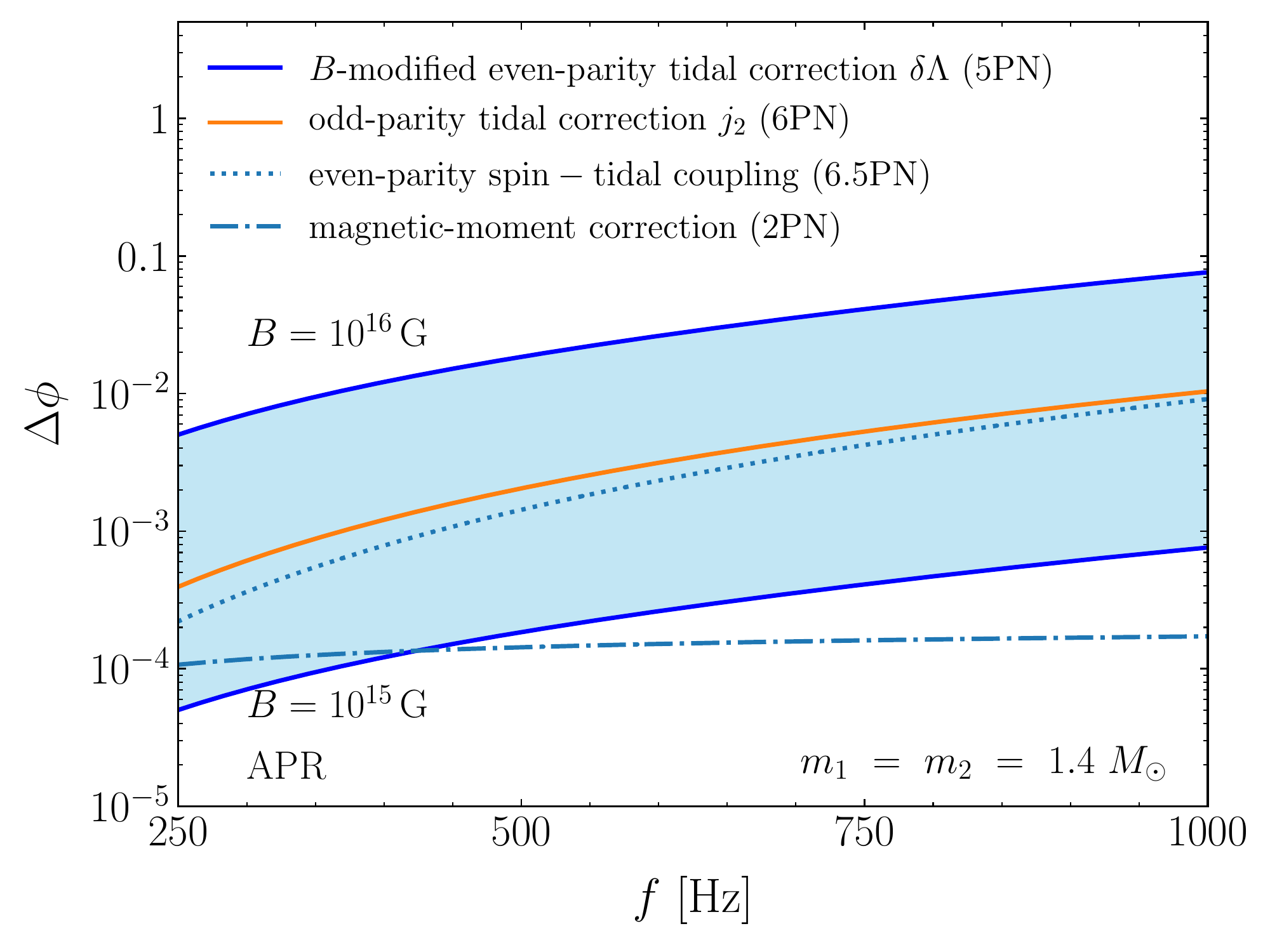}
  \includegraphics[width=0.49\textwidth]{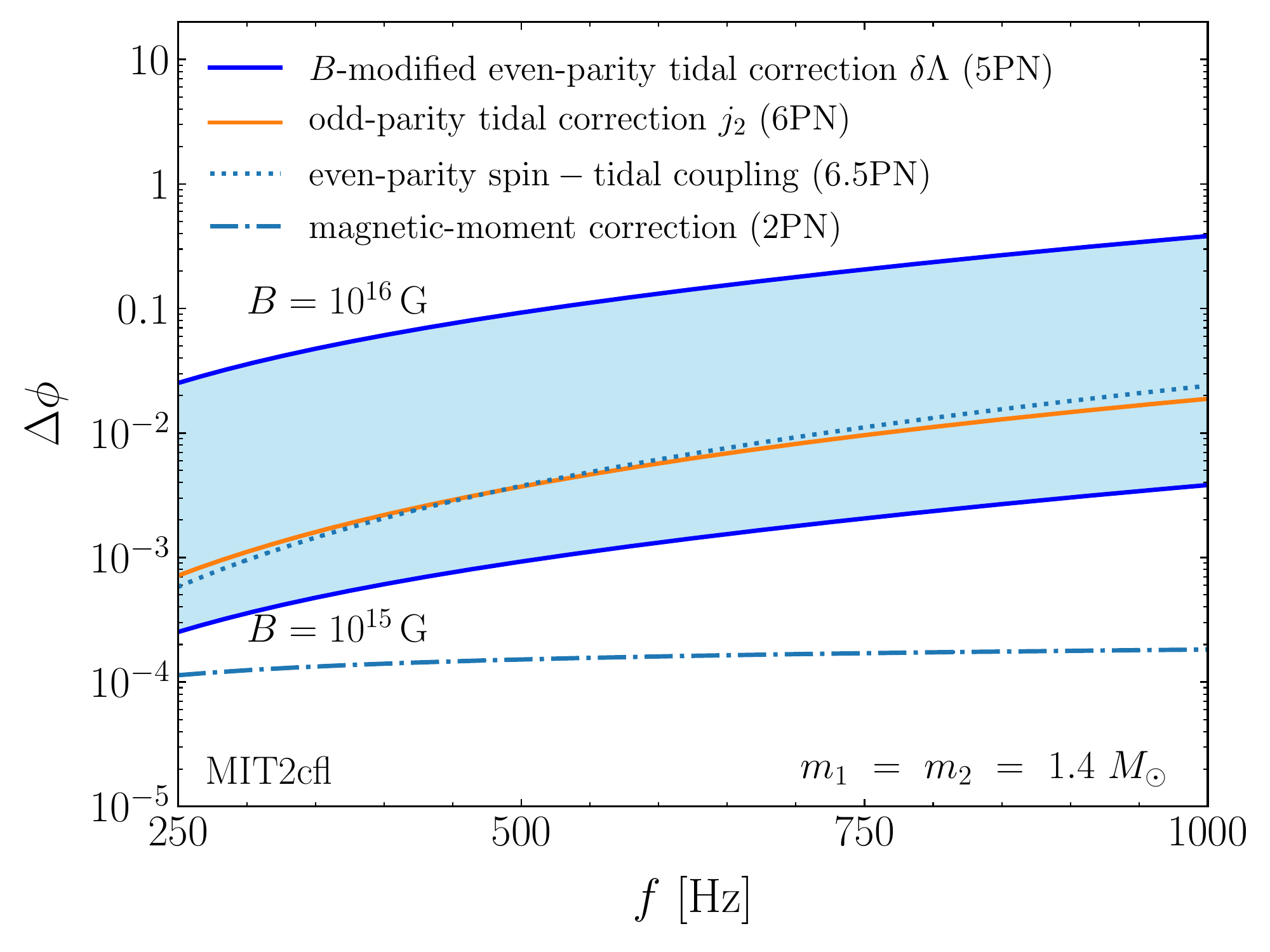}
  \caption{Evolution of the GW phase differences $\Delta \phi$ during the
    early inspiral (\ie for frequencies up to $1\,{\rm kHz}$). Shown as a
    comparison are the other the high-order non-magnetic corrections, \ie
    the odd-parity tidal correction $j_2$, the even-parity spin-tidal
    coupling, and the magnetic-moment correction. The blue-shaded region
    refers to magnetic-field strengths between $10^{15}\,{\rm G}$ and
    $10^{16}\,{\rm G}$. All the data refers to a representative binary
    with masses $m_1=m_2=1.4\,M_{\odot}$, with the left panel relative
    the APR EOS and the right one to the MIT2cfl EOS. }
\label{fig:phhigh}
\end{figure*}

All of the high-order corrections to the GW-phase evolution discussed
above are shown in Fig. \ref{fig:phhigh} as a function of the GW
frequency during the inspiral of an equal-mass binary with single mass
$M=1.4\ M_{\odot}$ and the APR EOS (left panel) or the MIT2cfl EOS (right
panel). Note that because the post-Newtonian approximation breaks down
near the merger, the phase difference is considered only up to a
frequency of $1000\,{\rm Hz}$. Shown in particular with a blue-shaded
region is the contribution of the magnetic tidal deformability with a
magnetic field strength from $10^{15}\, {\rm G}$ to $10^{16}\,{\rm
  G}$. Overall, Fig. \ref{fig:phhigh} shows that the contribution of
odd-parity tidal deformability and the even-parity spin-tidal corrections
for a dimensionless spin of $\chi=0.05$ (low-spin prior of GW170817) are
quite similar in size and frequency dependence. Both of them are larger
than the even-parity magnetic tidal deformability for $B=10^{15}\,{\rm
  G}$, but weaker than that for $B=10^{16}\,{\rm G}$.

Finally, we note that when the two stars are magnetised, the GW-phase
evolution during the inspiral is modified not only by pure gravitational
effects (\ie by the tidal deformation of the two stars), but also by the
loss of orbital energy to electromagnetic waves. The two stars, in fact,
can be assimilated to inspiralling dipoles that will generate
electromagnetic waves carrying away energy and angular momentum. The
corresponding correction to the binary dynamics appears at 2PN order and
was first computed by Ioka and Taniguchi \cite{Ioka2000}. Reported with
dot-dashed lines in Fig. \ref{fig:phhigh} is the strength of this
correction when calculated self-consistently with our magnetic-field
structure and for $B=10^{15}\,{\rm G}$. Clearly, this is the smallest of
the high-order contributions -- \ie between two and three orders of
magnitude smaller than the magnetically induced corrections to the tidal
deformability -- and grows only mildly with frequency, \ie as $f^{1/3}$.

In order to quantify the differences introduced by a magnetic field in
the GW waveforms of inspiralling binaries, we have computed the overlap
$\mathcal{O}$ between waveforms with or without magnetic field for
different EOSs and different detectors. We recall that the overlap is
defined as
\begin{eqnarray}
\mathcal{O} & := & \frac{\bras{h_{\delta \Lambda}}h_0\ket}{
\sqrt{\bras{h_{\delta \Lambda}} h_{\delta \Lambda \ket} 
\bras{h_0}h_0\ket}}\,, \label{eq:match1}
\end{eqnarray}
where the scalar product $\bras{h_{\delta \Lambda}}h_0\ket$ is given by
\begin{eqnarray}
\bras{h_{\delta \Lambda}}h_0\ket & := & \int_0^\infty 
\frac{\tilde{h}_{\delta \Lambda}(f) \tilde{h}_0^\ast (f)}{S_h(f)}df\,.
\label{eq:match2}
\end{eqnarray}
Here, $h_{\delta \Lambda}$ and $h_0$ represent the GW waveforms in the
time domain with and without $B$-modified tidal corrections, while
$\tilde{h}_{\delta \Lambda}$ and $\tilde{h}_0$ are the corresponding
Fourier transforms. Furthermore, since it is important to relate the
overlap with the actual sensitivity of a given detector, the quantity
$S_h(f)$ appearing in \eqref{eq:match2} is the noise power spectral
density of the detector under consideration, which in our analysis has
been considered for Advanced LIGO and ET.

In this way, we have found that across the various EOSs considered and
for a reference magnetic field $B=10^{15}\,{\rm G}$, the
\textit{mismatch}, \ie $\mathcal{M}:=1-\mathcal{O}$, is always extremely
small and of the order $\mathcal{M}\sim\ 10^{-9}$. These values are also
much smaller than the experimental limit for advanced LIGO, namely,
$\mathcal{O} \simeq 0.005$ \cite{Lindblom08,Giacomazzo:2009mp}; an
exception to this behaviour is offered by the quark-EOS ${\rm CIDDM}$,
which is the one leading to the largest phase difference. In this case,
and for an ultra-strong magnetic field $B=10^{16}\,{\rm G}$, we find the
mismatch to be $\mathcal{M}=0.003$, and thus slightly smaller than the
limit for LIGO.

Unfortunately, the use of a third-generation detector such as ET does not
help to increase the mismatch. This is because although the differences
in the waveforms obviously increase with a more sensitive detector that
will record a larger number of GW cycles, the total length of the
waveforms will also increase and so the normalisation in the denominator
of Eq. \eqref{eq:match1}. Fortunately, however, third-generation
detectors will also be able to have a finer determination of the tidal
deformability, \ie with a smaller experimental uncertainty. This was
considered in Ref. \cite{JimenezForteza2018}, where the posterior
distributions of the tidal deformability $\tilde{\Lambda}$ were computed
when considering the odd-parity tidal correction $j_2$. In that case, it
was shown that because of the high sensitivity of ET, the posterior
distributions of $\tilde{\Lambda}$ -- estimated when $j_2$ is computed
for either an irrotational or static fluid -- showed a significant
difference (see Fig. 6 in \cite{JimenezForteza2018}). Since we have shown
in Fig. \ref{fig:phhigh} that the odd-parity tidal correction $j_2$ is
actually smaller than the magnetic tidal deformability $\delta \Lambda$
when an extreme magnetic field of $B=10^{16}\,{\rm G}$ is considered, it
is possible that third-generation detectors would be able measure the
contributions to the phase evolution coming from the presence of
ultra-strong magnetic fields.

All things considered, we conclude that magnetically induced corrections
to the tidal deformability will determine changes in the GW-phase
evolution that are unlikely to be detected for realistic values of the
magnetic field (\ie $B\sim 10^{10}-10^{12}\,{\rm G}$), but that are
likely to produce a sizeable contribution should unrealistically large
magnetic fields (\ie $B\sim 10^{16}\,{\rm G}$) be present in the two
stars prior to merger.

\begin{figure}[!t]
\vspace{0.3cm}
{\centering
\resizebox*{0.48\textwidth}{!}
{\includegraphics{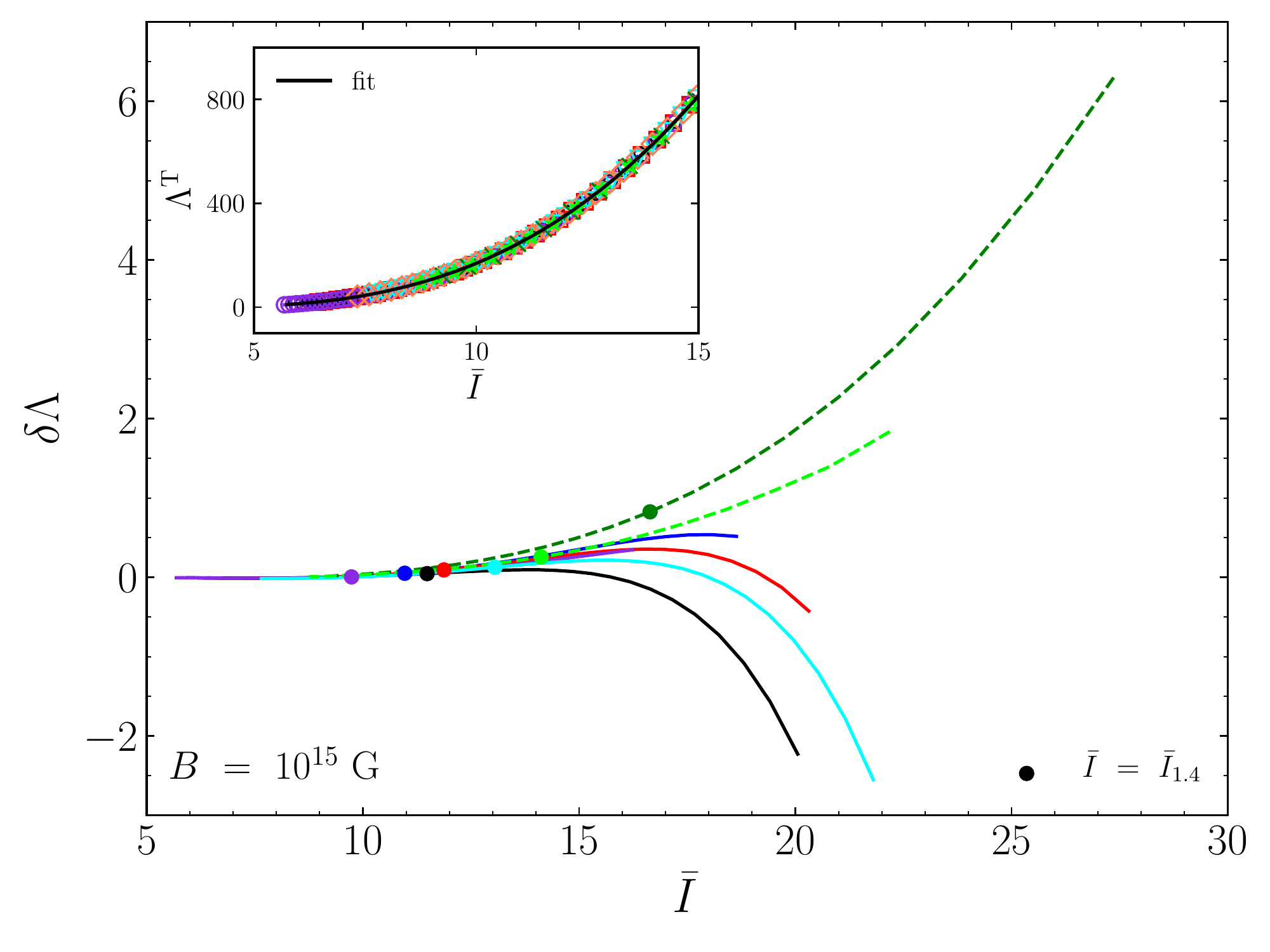}}
\par}
\caption{Broken universal relation between the magnetic tidal
  deformability and the dimensionless moment of inertia $\bar{I}:=I/M^3$
  for different EOSs and for a magnetic field of $B=10^{15}\, {\rm G}$
  (different magnetic-field strengths will only rescale the vertical axis
  but do not change the functional behaviour). The loss of universality
  between $\delta \Lambda$ and $\bar{I}$ does not impact significantly
  the overall universality between $\Lambda$ and $\bar{I}$ (see inset).
}\label{fig:ilove}
\end{figure}

\subsection{On the validity of universal relations}
\label{sec:IIID}

The last topic we will discuss briefly here is the issue of the validity
of the quasi-universal relations that have been shown to exist between
the moment of inertia, the Love number, and the mass quadrupole of
nonrotating compact stars \cite{Yagi2013a}. While these relations have
been demonstrated to hold under a very broad set of conditions (see
\cite{Yagi2017, Doneva2018} for some recent reviews), they have also been
shown to be lost in the case of strong magnetic fields \cite{Haskell2014},
or to be modified during the inspiral \cite{Maselli2013}. Since such
strong magnetic fields are often invoked in our analysis of the tidal
deformability, it is reasonable to consider under what conditions the
universal relations between the magnetic tidal deformability $\delta
\Lambda$ and the moment of inertia $I$ break-down when considering the
poloidal magnetic-field configurations explored here.

We note that a somewhat similar analysis was carried out in
Ref. \cite{Haskell2014}, which was however focused on the validity of the
universal relation between the moment of inertia $I$ and the stellar
quadrupole moment $Q$ when considering a twisted-torus magnetic topology
\cite{Ciolfi2009,Ciolfi2013}. In that work, it was found that for simple
magnetic-field configurations that are purely poloidal or purely
toroidal, the relation between the stellar quadrupole moment $Q$ and the
moment of inertia $I$ is nearly universal. However, different magnetic
field geometries lead to different $I$–$Q$ relations, and, in the case of
a twisted-torus configuration, the relation depends significantly on the
EOS, losing its universality. In particular, universality was found to be
lost for stars with long spin periods, \ie $P\gtrsim 10\, {\rm s}$, and
strong magnetic fields, \ie $B\gtrsim 10^{12}\, {\rm G}$.

The results of the analysis for the $I$-$\Lambda$ universal relation is
summarised in Fig. \ref{fig:ilove}, which reports in the main panel
$\delta \Lambda$ as a function of the dimensionless moment of inertia
$\bar{I}:=I/M^3$ for different EOSs and for a magnetic field of
$B=10^{15}\, {\rm G}$ (different magnetic-field strengths will only
change the vertical scale, but not the functional behaviour); marked with
filled circles are the values for $1.4\,M_{\odot}$ stars. What can be
easily appreciated from the main panel in Fig. \ref{fig:ilove} is that no
universal relation can be found between $\delta \Lambda$ and $\bar{I}$
and that the curves relative to different EOSs deviate form a universal
behaviour for $\bar{I} \gtrsim 10$. Furthermore, quark stars and neutron
stars show a distinctively different behaviour, with $\delta \Lambda$
increasing monotonically with $\bar{I}$, while decreasing for neutron
stars. Indeed, we have already encountered a similar behaviour in
Fig. \ref{fig:eos} and this does not come as a surprise since $\bar{I}
\sim (R/M)^2$. As a final remark, we note that the loss of universality
between $\delta \Lambda$ and $\bar{I}$ does not impact significantly the
overall universality between $\Lambda$ and $\bar{I}$, which is preserved
by the fact that $\delta \Lambda \ll \Lambda^{\rm T}$, and that
$\Lambda^{\rm T}$ still correlates universally with $\bar{I}$ for
$\bar{I} \lesssim 15$, as shown in the inset in Fig. \ref{fig:ilove}.

\section{Conclusions}
\label{sec:summary}

The evolution of the GW phase produced by inspiralling binaries of
compact stars is subject to corrections coming from the nonzero
deformability of the two stars. In turn, because the tidal deformability
is directly related to the properties of the EOS of nuclear matter, the
measurement of these corrections promises to be an important tool to
read-off the EOS from the GW signal. Extensive work has been carried out
over the last decade to quantify in an even more accurate manner the size
of these corrections when taking into account a number low- and
high-order corrections to the tidal deformability coming, for instance,
by mass-current multipoles or by the presence of an intrinsic spin in the
star. This bulk of work has reached a considerable level of
sophistication and a rather comprehensive view of this problem is now
available in the literature.

We have here considered an aspect of this research that has not been
explored so far, namely, the high-order corrections to the tidal
deformability that are introduced by the presence of an intrinsic
magnetic field in the stars. These corrections should not be confused
with the ``gravitomagnetic'' (or odd-parity) corrections to the tidal
deformability, namely, with the ``even-parity'' quadrupolar tidal
deformability, which starts to impact the phase of the GW signal at 5PN.

At the order considered here, the magnetic field induces correction only
to the even-parity quadrupole moment and we assume that it does not lead
to coupling of different multipole moments. However, because these
represent a correction to the standard unmagnetised, nonspinning tidal
deformability, they impose an analysis that includes second-order
perturbations. Proceeding in this way, we were able to compute the
magnetic-field induced changes to the tidal deformability and to assess
their impact on the evolution of the GW phase for different strengths of
the magnetic field and for different EOSs, including those that describe
quark stars. Overall, we find that magnetically induced corrections to
the tidal deformability will produce changes in the GW-phase evolution
that are unlikely to be detected if the magnetic field has the strength
expected from astrophysical considerations, \ie $B\sim 10^{10} -
10^{12}\,{\rm G}$. At the same time, if the magnetic field present in the
two stars prior to merger is unrealistically large, \ie $B\sim
10^{16}\,{\rm G}$, these corrections are expected to produce a sizeable
contribution to the GW-phase evolution measured by third-generation
detectors such as ET and CE. In this unlikely event, and if the
  neglected higher-order terms will remain negligible also for very high
  magnetic fields, the induced phase differences would represent a very
useful tool to study and measure the properties of the magnetic fields in
the merging stars, thus providing information that is otherwise hard to
quantify.

\begin{acknowledgments}

We thank Valeria Ferrari, Tanja Hinderer, Ian Harry and Michail Chabanov
for valuable discussions. Support comes in part from ``PHAROS'', COST
Action CA16214, the LOEWE- Program in HIC for FAIR and Natural Science
Foundation of China (No. 11873040).
\end{acknowledgments}

\appendix
  
\section{Details on the derivation}
\label{sec:appendix_A}

In what follows we provide details on the derivation of the perturbative
equations presented in the main text and that were omitted from
compactness. We start by recalling that the the nonvanishing components
of the perturbed Einstein tensor are $\delta G_{\ 0}^0$, $\delta
G_{\ i}^i$ and $\delta G_{\ 1}^2$. In particular, the Einstein equation
$\delta G_{\ 2}^2 - \delta G_{\ 3}^3 = 0$ can be used to obtain a
relation between $\delta H_0$ and $\delta H_2$ as
\begin{align}\label{eq:gra23}
\delta H_0 + \delta H_2 = A_{23}\,, 
\end{align}
where $A_{23}$ is an additional term resulting from the first-order
perturbation
\begin{align}\label{eq:a23}
A_{23} :=  &-\frac{4}{7} \biggl[H_0 h_2^{\rm B} - \frac{H_0 m_2^{\rm B} e^{\lambda}}{r} \biggr] + \biggl[ 2h_0^{\rm B} H_0 - \frac{2H_0 m_0^{\rm B} e^{\lambda}}{r} \biggr]\nonumber \\
&  - \frac{4}{21} \biggl[3K h_2^{\rm B} + 2K e^{-\lambda}(a_1^{\prime})^{2} + 3H_0 h_2^{\rm B} + \frac{3K m_2^{\rm B} e^{\lambda}}{r} - \frac{3H_0 m_2^{\rm B} e^{\lambda}}{r} \biggr] \nonumber \\
&  + \frac{6}{7} \biggl[ H_0 h_2^{\rm B} + \frac{K m_2^{\rm B} e^{\lambda}}{r} - \frac{H_0 m_2^{\rm B} e^{\lambda}}{r} + K h_2^{\rm B} \biggr]\,. 
\end{align}

Similarly, the Einstein equation $\delta G_{\ 1}^2 = 8\pi\delta T_{\ 1}^2
= 0$ can be exploited to relate $\delta H_0$ and $\delta K^\prime$ as
\begin{align}\label{eq:gra12}
\frac{r^2 \nu^\prime}{4} \biggl(\delta H_0^{\ell} - \delta H_2^{\ell} \biggr) + \frac{r^2 }{2} \biggl(\delta K_{\ell}^\prime + \delta H_0^{l\prime} \biggr) - \frac{r}{2} \biggl(\delta H_0^{\ell} + \delta H_2^{\ell} \biggr) = B_{12}\,,  
\end{align}
where $B_{12}$ is defined as
\begin{align}\label{eq:b12}
B_{12} := &\frac{2}{7}r^2 K k_2^{{\rm B}\prime} + \frac{1}{7} \left[ \frac{r^2\nu^\prime}{2} H_0 h_2^{\rm B}  +  r^2 H_0^{\prime} h_2^{\rm B} + r^2 K^\prime k_2^{\rm B} - r H_0 h_2^{\rm B} + \frac{r \nu^\prime H_0 m_2^{\rm B} e^{\lambda} }{2} \right] \nonumber \\
& + \left[\frac{r^2 \nu^\prime}{2} H_0 h_0^{\rm B}  + r^2 H_0^{\prime} h_0^{\rm B} + \frac{r \nu^\prime H_0 m_0^{\rm B} e^{\lambda}}{2} - r H_0 h_0^{\rm B} + H_0 m_0^{\rm B} e^{\lambda} \right] \nonumber \\
&  - \frac{1}{7}\biggl[r^2 H_0 k_2^{{\rm B}\prime} + \frac{r^2 \nu^\prime H_0 h_2^{\rm B}}{2} - r^2 K^\prime k_2^{\rm B}  - \frac{r \nu^\prime H_0 m_2^{\rm B} e^{\lambda}}{2}\biggr] \nonumber \\
&  + \frac{4}{21} K a_1 a_1^\prime +  \frac{3}{7} H_0 m_2^{\rm B} e^{\lambda}\,. 
\end{align}

The remaining Einstein equations are
\begin{align}
  \label{eq:g00}
  &\delta G_{\ 0}^0 - \delta G_{\ 1}^1 = -8\pi \left(1 + \frac{1}{c_s^2}\right)\,\delta p^{\rm BT}\,,\\
  \label{eq:g22}
  &\delta G_{\ 2}^2 - \delta G_{\ 3}^3 = 16\pi \delta p^{\rm BT}\,,
\end{align}
where $\delta p^{\rm BT}$ is the second-order pressure perturbation. We
can further define two functions, $C_{01}$ and $C_{23}$, to simplify
Eqs. \eqref{eq:g00}, \eqref{eq:g22} as 
\begin{align}\label{eq:c01}
C_{01} := & \frac{2}{7} \biggl[K \biggl(\lambda^\prime k_2^{{\rm B}\prime} + \nu^\prime k_2^{{\rm B}\prime} - 2 k_2^{{\rm B}\prime\prime} \biggr)e^{-\lambda} - H_0 \biggl(\lambda^\prime k_2^{{\rm B}\prime} + \nu^\prime k_2^{{\rm B}\prime} - 2 k_2^{{\rm B}\prime\prime} \biggr) e^{-\lambda} - K^\prime \biggl(h_2^{{\rm B}\prime} + 2 k_2^{{\rm B}\prime} \biggr) e^{-\lambda} \nonumber \\
& + \biggl( K^\prime \lambda^\prime + K^\prime \nu^\prime - 2 K^{\prime\prime} \biggr) k_2^{\rm B} e^{-\lambda}   - 4\biggl(K - H_0 \biggr)\frac{ k_2^{{\rm B}\prime} e^{-\lambda}}{r}  + \frac{ 2 H_0^{\prime} h_2^{\rm B} e^{-\lambda}}{r} -\frac{ K^\prime m_2^{{\rm B}\prime}}{r} - \frac{ 6K h_2^{\rm B} }{r^2} \nonumber \\
& -\frac{ 4 K^\prime k_2^{\rm B} e^{-\lambda}}{r} + \biggl(\nu^\prime K^\prime - 2 K^{\prime\prime} \biggr)\frac{ m_2^{\rm B}}{r}  - \frac{ 6H_0 h_2^{\rm B}}{r^2}  - \frac{ 12 H_0 k_2^{\rm B}}{r^2} + \frac{ 4 H_0}{r^2}\biggl(\nu^\prime m_2^{\rm B} - m_2^{{\rm B}\prime} \biggr)  \nonumber \\
& - \biggl( 3K^\prime + 2 H_0^{\prime} \biggr) \frac{m_2^{\rm B}}{r^2} - \frac{ 6H_0 h_2^{\rm B}}{r^2} + \frac{ 6K m_2^{\rm B} e^{\lambda}}{r^3} + \frac{ H_0 m_2^{\rm B} e^{\lambda}}{r^3} \biggl(4e^{-\lambda} - 6 \biggr) - \frac{ 6H_0 m_2^{\rm B} e^{\lambda}}{r^3}  \biggr] \nonumber \\
& + \biggl[ \frac{ 2 H_0^{\prime} h_0^m e^{-\lambda}}{r} + \biggl(\nu^\prime K^\prime - 2 K^{\prime\prime} \biggr) \frac{ m_0^{\rm B}}{r}  - \frac{K^\prime m_0^{{\rm B}\prime}}{r} -  K^\prime h_0^{{\rm B}\prime} e^{-\lambda} - \frac{6 H_0 h_0^{\rm B} }{r^2} - \biggl( 3 K^\prime + 2 H_0^{\prime} \biggr) \frac{m_0^{\rm B}}{r^2} \nonumber \\
&  + \frac{ 4 H_0}{r^2} \biggl(\nu^\prime m_0^{\rm B} - m_0^{{\rm B}\prime} \biggr) + \frac{H_0 m_0^{\rm B} e^{\lambda}}{r^3} \biggl(4e^{-\lambda} - 6 \biggr)  \biggr] 
 + \frac{6}{7} \biggl[ \frac{ H_0 h_2^{\rm B}}{r^2} + \frac{ H_0 m_2^{\rm B} e^{\lambda} }{r^3} \biggr]  + \frac{88}{21} \frac{ H_0 a_1^2 }{r^4}\,, 
\end{align}
\begin{align}\label{eq:c23}
C_{23} := & \frac{2}{7}\biggl[ K \biggl(\lambda^\prime k_2^{{\rm B}\prime} - \nu^\prime k_2^{{\rm B}\prime} - 2 k_2^{{\rm B}\prime\prime} \biggr)e^{-\lambda} - H_0 \biggl(\lambda^\prime k_2^{{\rm B}\prime} - \nu^\prime k_2^{{\rm B}\prime} - 2 k_2^{{\rm B}\prime\prime} \biggr) e^{-\lambda} + \biggl( \lambda^\prime H_0^{\prime} - 2 \nu^\prime H_0^{\prime} - 2 H_0^{\prime\prime} \biggr) e^{-\lambda}h_2^{\rm B}  \nonumber \\
& - \frac{64 \pi p H_0 m_2^{\rm B} e^{\lambda}}{r} + \biggl(\lambda^\prime K^\prime - \nu^\prime K^\prime  - 2 K^{\prime\prime} \biggr)k_2^{\rm B} e^{-\lambda}  + \biggl(K^\prime - H_0^{\prime} \biggr) \biggl(h_2^{{\rm B}\prime} - 2 k_2^{{\rm B}\prime} \biggr) e^{-\lambda}  \nonumber \\
& - \frac{ 4 K k_2^{{\rm B}\prime} e^{-\lambda}}{r} + \frac{ 4 H_0 k_2^{{\rm B}\prime} e^{-\lambda}}{r} - \frac{2 \nu^\prime}{r} H_0 \biggl(\lambda^\prime m_2^{\rm B} + m_2^{{\rm B}\prime}\biggr) - \frac{2 H_0^{\prime} h_2^{\rm B} e^{-\lambda} }{r} - \frac{ 4 K^\prime k_2^{\rm B} e^{-\lambda}}{r} \nonumber \\
& - \biggl(K^\prime \nu^\prime + 2 K^{\prime\prime} + 4 H_0^{\prime} \nu^\prime + 2 H_0^{\prime\prime} \biggr)\frac{ m_2^{\rm B}}{r} - \biggl(K^\prime + H_0^{\prime} \biggr) \frac{ m_2^{{\rm B}\prime}}{r} + \frac{ 6K h_2^{\rm B}}{r^2} + \frac{ 6 H_0 h_2^{\rm B}}{r^2} \nonumber \\
& - \frac{ 2 H_0}{r^2} \biggl(2 \lambda^\prime m_2^{\rm B} - \nu^\prime m_2^{\rm B} + 2 m_2^{{\rm B}\prime} \biggr) - \biggl(3 K^\prime + 5 H_0^{\prime} \biggr) \frac{ m_2^{\rm B}}{r^2} + \frac{ 6H_0 h_2^{\rm B}}{r^2} \nonumber \\
& + \frac{ 6K m_2^{\rm B} e^{\lambda}}{r^3} - \frac{ 6H_0 m_2^{\rm B} e^{\lambda}}{r^3} + \frac{ 4 H_0 m_2^{\rm B} }{r^3} - \frac{ 6H_0 m_2^{\rm B} e^{\lambda}}{r^3} \biggr]  \nonumber  \\
& \biggl[ - \frac{64 \pi p H_0 m_0^{\rm B} e^{\lambda}}{r} + \biggl(\lambda^\prime H_0^{\prime} - 2 \nu^\prime H_0^{\prime}  - 2 H_0^{\prime\prime} \biggr) h_0^{\rm B} e^{-\lambda} +  \left(K^\prime- H_0^{\prime}\right) h_0^{{\rm B}\prime} e^{-\lambda}  - \frac{ 2 H_0^{\prime} h_0^{\rm B} e^{-\lambda} }{r}  \nonumber \\
& -\frac{2 \nu^\prime H_0 }{r}\biggl(\lambda^\prime m_0^{\rm B} + m_0^{{\rm B}\prime} \biggr) - \biggl(\nu^\prime K^\prime + 2 K^{\prime\prime} + 4\nu^\prime H_0^{\prime} + 2 H_0^{\prime\prime} \biggr) \frac{ m_0^{\rm B}}{r} - \biggl(K^\prime + H_0^{\prime} \biggr)\frac{ m_0^{{\rm B}\prime}}{r} \nonumber \\
& + \frac{ 6H_0 h_0^{\rm B} }{r^2} - \frac{ 2 H_0 }{r^2} \biggl( 2 \lambda^\prime m_0^{\rm B} - \nu^\prime m_0^{\rm B} +  2 m_0^{{\rm B}\prime} \biggr) - \biggl(3 K^\prime + 5 H_0^{l\prime} \biggr)\frac{ m_0^{\rm B} }{r^2} + \frac{ 4 H_0 m_0^{\rm B} }{r^3} - \frac{ 6H_0 m_0^{\rm B} e^{\lambda}}{r^3} \biggr] \nonumber \\
& + \frac{6}{7} \biggl[\frac{ 2 H_0 m_2^{\rm B} e^{\lambda} }{r^3} - \frac{ 2 H_0 h_2^{\rm B} }{r^2} \biggr] - \frac{88}{21}\frac{ K a_1^2 }{r^4}\,. 
\end{align}

The final form that Eqs. \eqref{eq:g00} and \eqref{eq:g22} then take is 
\begin{align}\label{eq:gra01}
C_{01}  = & -8\pi(1 + \frac{1}{c_s^2})\delta \tilde{P} - \frac{ 3(\delta H_0 - \delta H_2)}{r^2} + \frac{ e^{-\lambda} }{2} \biggl(\lambda^\prime \delta K^\prime + \nu^\prime \delta K^\prime - 2\delta K^{\prime\prime} \biggr) \nonumber \\
& -\frac{ \delta H_2 e^{-\lambda} (\lambda^\prime + \nu^\prime)}{r} - \frac{e^{-\lambda}}{r} \biggl(2 \delta K^\prime - \delta H_0^{\prime} - \delta H_2^{l\prime} \biggr)\,, 
\end{align}
\begin{align}\label{eq:grac23}
C_{23} = & 16\pi \delta \tilde{P} + 16 \pi p \delta H_2 + \frac{ e^{-\lambda} \delta K^\prime (\lambda^\prime - \nu^\prime)}{2} + \frac{ e^{-\lambda}}{2} \biggl(\lambda^\prime \delta H_0^{\prime} + \nu^\prime \delta H_2^{\prime} \biggr) \nonumber \\
&  - e^{-\lambda} \biggl(\delta K^{\prime\prime} + \delta H_0^{\prime} \nu^\prime + \delta H_0^{\prime\prime} \biggr) - \frac{ e^{-\lambda}}{r} \biggl(2 \delta K^\prime + \delta H_0^{\prime} - \delta H_2^{\prime} \biggr) + \frac{ 3(\delta H_0 + \delta H_2)}{r^2}\, 
\end{align}
where $\delta \tilde{P}$ is defined as
\begin{align}\label{eq:dpt}
\delta \tilde{P} & := \frac{5}{2}\int_0^\pi \delta p^{\rm BT} P_2(\cos\theta)\sin\theta d\theta\,. 
\end{align}

Finally, we can substitute in Eq. \eqref{eq:gra01} the expressions for
$\delta H_2$, $\delta K$, and $\delta \tilde{P}$ given by
Eqs.~\eqref{eq:gra23}--\eqref{eq:gra12}, \eqref{eq:grac23}. In this
way, we obtain the master equation for $\delta H_0$, \ie
Eq.~\eqref{eq:eqmix}, whose source term is explicitly given by
\begin{align}\label{eq:sour}
S(r) & = \biggl\{2C_{01} + (1+\frac{1}{c_s^2})C_{23} - e^{-\lambda}(\lambda^\prime + \nu^\prime)T + 2e^{-\lambda}T^\prime + \frac{4e^{-\lambda}}{r}T + \frac{6}{r^2}A_{23} + \frac{2A_{23}e^{-\lambda}(\lambda^\prime + \nu^\prime)}{r} - \frac{2e^{-\lambda}}{r}A_{23}^\prime \nonumber \\
& - \left[16\pi pA_{23} + \frac{e^{-\lambda}}{2}A_{23}^\prime\nu^\prime + \frac{e^{-\lambda}}{r}A_{23}^\prime + \frac{3}{r^2}A_{23} + \frac{T}{2}e^{-\lambda}(\lambda^\prime - \nu^\prime) - e^{-\lambda}T^\prime - \frac{2e^{-\lambda}}{r}T \right](1+\frac{1}{c_s^2}) \biggr\}\frac{e^\lambda}{2}\,, 
\end{align}
where we have defined
\begin{align}\label{eq:t}
T & := (r\nu^\prime + 2)\frac{A_{23}}{2r} + \frac{2B_{12}}{r^2}. 
\end{align}

The exterior source term is obtained readily after setting to zero the
matter terms of Eq.~\eqref{eq:sour} and it is therefore given by
\begin{align}\label{eq:sour_ext}
S^e(z) = & \frac{9c_1^e \mu^2}{224 M^4}(3z^4 + 176z^3 - 180z^2 - 140z -147) \log\left(\frac{z+1}{z-1} \right)^3  \nonumber \\
& + \biggl[ \frac{3c_2^e\mu^2}{112M^4}(3z^4 + 176z^3 - 180z^2 - 140z -147) + \frac{18K^{\rm B} c_1^e}{7} (4z^4 + 15z^3 - 2z^2 - 15z - 2 )  \nonumber \\
& + \frac{3c_1^e\mu^2}{112M^4} \frac{72z^6 + 243z^5 - 936z^4 - 1566z^3 + 1216z^2 + 723z - 488}{z^2-1}  \biggr]\log\left(\frac{z+1}{z-1} \right)^2 \nonumber \\
& + \biggl[ \frac{3c_2^e\mu^2}{14M^4}\frac{6z^6 + 21z^5 - 34z^4 - 86z^3 + 37z^2 + 9z - 37}{z^2-1}   + \frac{12c_1^e c_2}{M} \frac{-z^2 + z + 2}{z-1} \nonumber \\
&  + \frac{3c_1^e\mu^2}{56M^4}\frac{-144z^9 - 513z^8 + 528z^7 + 2385z^6 - 500z^5 - 2119z^4 + 564z^3 - 621z^2 - 640z + 772}{z^6 - 3z^4 + 3z^2 - 1 } \nonumber \\
& + \frac{12K^{\rm B}c_2^e}{7}(4z^4 + 15z^3 - 2z^2 - 15z - 2) - \frac{24K^{\rm B}c_1^e}{7}\frac{12z^5 + 45z^4 - 14z^3 - 75z^2 + 6z + 12}{z^2-1} \biggr] \log\left(\frac{z+1}{z-1} \right) \nonumber \\
& + \biggl[-\frac{3c_2^e\mu^2}{28M^4}\frac{24z^7 + 87z^6 + 24z^5 - 224z^4 - 216z^3 - 23z^2 + 132z + 228}{z^4 - 2z^2 + 1} + \frac{8c_2^e c_2}{M} \frac{-z^2 + z + 2}{z - 1} \nonumber \\
& + \frac{3c_1^e \mu^2}{28M^4} \frac{72z^8 + 261z^7 + 24z^6 - 846z^5 - 648z^4 + 85z^3 + 456z^2 + 764z - 144}{z^6 - 3z^4 + 3z^2 - 1} \nonumber \\
& + \frac{8c_1^e c_2}{M} \frac{3z^3 - 3z^2 - 8z + 2}{z^3 - z^2 - z + 1} - \frac{8c_2^e K^{\rm B}}{7} \frac{12z^5 + 45z^4 - 14z^3 - 75z^2 + 6z + 12}{z^2 - 1} \nonumber \\
& + \frac{8c_1^e K^{\rm B}}{7} \frac{36z^8 + 135z^7 - 102z^6 - 450z^5 + 136z^4 + 447z^3 - 110z^2 - 120z + 64}{z^6 - 3z^4 + 3z^2 - 1} \biggr]. 
\end{align}

We conclude by reporting the contribution from surface source term
$S_{\rm surf}$ in the matching procedure of Eqs. \eqref{eq:matching2q1} 
  and \eqref{eq:matching2q2}, namely,
\begin{align}\label{eq:sour_surf}
S_{\rm surf}(r) = & \biggl(C_{23} - 16\pi pA_{23} - \frac{e^{-\lambda}}{2}A_{23}^\prime\nu^\prime - \frac{e^{-\lambda}}{r}A_{23}^\prime - \frac{3}{r^2}A_{23} - \frac{T}{2}e^{-\lambda}(\lambda^\prime - \nu^\prime) + e^{-\lambda}T^\prime + \frac{2e^{-\lambda}}{r}T \biggr)\biggl|_{r=R} \frac{R^2}{2M}\,. 
\end{align}

\bibliography{aeireferences}

\end{document}